\documentclass[11pt]{article}
\usepackage{graphicx}
\usepackage{enumerate}
\usepackage{amsmath}
\usepackage{amsthm}
\usepackage{amsfonts}
\usepackage[dvipsnames]{xcolor}
\usepackage[
  bookmarks=false,
  pdfpagelabels=false,
  hyperfootnotes=false,
  hyperindex=false,
  pageanchor=false,
  colorlinks,
]{hyperref}
\usepackage{setspace, amssymb}

\newtheoremstyle{dotless}{}{}{}{}{\bfseries}{}{ }{}
\theoremstyle{dotless}
\newtheorem{thm}{Theorem}
\newtheorem{rem}{Remark}
\newtheorem{defn}{Definition}
\newtheorem{res}{Result}
\newtheorem{lem}{Lemma}
\newtheorem{prop}{Proposition}
\newtheorem{cor}{Corollary}
\newtheorem{ob}{Observation}
\newtheorem{ex}{Example}

\newtheorem{ax}{Axiom}
\newtheorem{innernewax}{Axiom}
\newenvironment{newax}[1]
  {\renewcommand\theinnernewax{#1}\innernewax}
  {\endinnernewax}

\newtheorem{innernewlem}{Lemma}
\newenvironment{newlem}[1]
  {\renewcommand\theinnernewlem{#1}\innernewlem}
  {\endinnernewlem}

\newcommand{\emp}{\varnothing}
\newcommand{\bt}{\begin{theorem}}
\newcommand{\wt}{\widetilde}

\newcommand{\td}{\hspace{0.1cm}\hat{\neq}\hspace{0.1cm}}
\newcommand{\sm}{\setminus}

\newcommand{\et}{\end{theorem}}
\newcommand {\be}{\begin{equation}}
\newcommand {\ee}{\end{equation}}

\newcommand{\hsp}{\hskip 2em}
\newcommand{\ul}{\underline}

\newcommand{\ve}{\varepsilon}
\newcommand{\ol}{\overline}
\newcommand{\noi}{\noindent}

\def \qed {\hfill \vrule height6pt width6pt depth0pt}

\title{A characterization of lexicographic preferences}

\date{\today}

\begin{document}
\author{\textsc{Mridu Prabal Goswami}{\footnote{Indian Statistical Institute, Tezpur, India}}\hsp
\textsc{Manipushpak Mitra}{\footnote{Indian Statistical Institute, Kolkata, India}}\hsp
\textsc{Debapriya Sen}{\footnote{Ryerson University, Toronto, Canada\vspace{0.2cm}}}}
\date{July 31, 2021}
\maketitle

\maketitle
\onehalfspacing

\begin{abstract}\noi This paper characterizes lexicographic preferences over alternatives that are identified by a finite number of attributes. Our characterization is based on two key concepts: a weaker notion of continuity called `mild continuity' (strict preference order between any two alternatives that are different with respect to every attribute is preserved around their small neighborhoods) and an `unhappy set' (any alternative outside such a set is preferred to all alternatives inside). Three key aspects of our characterization are: (i) use of  continuity arguments, (ii) the stepwise approach of looking at two attributes at a time and (iii) in contrast with the previous literature, we do not impose noncompensation on the preference and consider an alternative weaker condition.

\end{abstract}

\noindent
{\bf Keywords:} lexicographic preferences; mild continuity; induced preferences; unhappy set; inclusion of marginally improved alternatives (IMIA)

\newpage

\section{Introduction}

Consider an individual decision problem over alternatives that have multiple attributes. For an individual having a lexicographic preference over such alternatives, there is an order of importance of different attributes. If two alternatives differ with respect to the most important attribute, the individual prefers the one that is superior in regard to that attribute. If two alternatives are same with respect to the most important attribute, he prefers the one that is superior in regard to the second most important attribute and so on. As put succintly by Bettman et al. (1998, p.190):
\begin{quote}
``The lexicographic strategy involves limited, attribute-based, noncompensatory processing that is selective across attributes and consistent across alternatives."
\end{quote}
The decision process under a lexicographic preference has two salient features. First, as the individual looks at one attribute at a time, the process is stepwise or sequential in nature. Second, it is noncompensatory in that it does not allow for any tradeoffs between attributes: superiority in a less important attribute does not compensate for the inferiority in a more important attribute. While the stepwise nature of the decision process is appealing particularly for problems with a large number of attributes, the noncompensatory property imposes a kind of rigidity on decision making that may not be  reasonable (see, e.g., Ford et al. 1989, Keeney and Raiffa 1993, Kurz-Milcke and Gigerenzer 2007).  For this reason, a lexicographic preference is often considered to be more useful as a heuristic  rather than an exact decision rule.\footnote{According to Kurz-Milcke and Gigerenzer (2007): ``The study of heuristics analyzes how people make decisions when optimization is out of reach." Leong and Hensher (2012) point out the evidence from the psychology literature that suggests ``humans rely on the use of quick mental processing rules known as decision heuristics to manage the vast number of decisions". See also Goldstein and Gigerenzer (1999), Hogarth and Karelaia (2006), Yee et al. (2007), Gigerenzer and Gaissmaier (2011) and Katsikopoulos (2011, 2013) for an overview of models and experiments of lexicographic and other related decision rules such as recognition and take-the-best heuristics.} 

On closer look, the noncompensatory property is unreasonable mainly because what it prescribes at the margin. If alternative $a$ is only marginally inferior to alternative $b$ in regard to the most important attribute but much superior in regard to the second most important attribute, the lexicographic preference still demands that $b$ is preferred to $a.$ Tversky (1969) addressed this drawback by proposing the lexicographic semiorder (LS), which prescribes that $b$ is  preferred to $a$ if $b$ is superior to $a$ in regard to the most important attribute by a magnitude that exceeds a specific threshold $\ve;$ but if this magnitude does not exceed $\ve$ and $a$ is superior in regard to the less important attribute, then $a$ is preferred. Using LS for experiments with gambles where any gamble has two attributes, its probability of winning the prize and its prize money, Tversky (1969) found evidence of intransitive preference behavior, although more recent  experimental literature does not find conclusive support for intransitivity (see, e.g., Gonz\'{a}lez-Vallejo 2002, Birnbaum and Gutierrez 2007). 

The noncompensatory property in itself is not problematic when differences are not marginal for an individual decision maker. This is actually the case in certain  real life decision problems where the available alternatives are such that when they differ with respect to an attribute, the difference is not negligible. If a decision maker does not face comparisons involving marginal differences, the unreasonable rigidity of the noncompensatory property is not an issue and there is merit in applying the lexicographic decision rule. It should be noted in this regard that marginal differences can be subjective; a difference that is considered marginal by an individual may not be viewed as such by another individual.

A smartphone is a product with multiple attributes. Two models of a smartphone can be same in regard to some  attributes, but if they differ with respect to an attribute, the difference is usually not marginal for many buyers. For instance, Galaxy A51 and Galaxy A71 models of Samsung\footnote{For detailed configurations of these two and other models, see 
\url{https://www.samsung.com/ca/smartphones/all-smartphones/}}  in Canada compare as follows in regard to three key attributes: price, storage capacity and camera quality. Galaxy A51 has price $\$479.99,$ internal storage 64GB and main camera resolution 48MP, while the corresponding values for Galaxy A71 are $\$599.99,$ 128GB and 64MP. Choosing between these two models, many buyers do not face a situation of marginal differences such as one model offering a superior camera quality at a price that is only slightly higher than the other model.  

Similar examples can be found in the banking sector where a product (e.g., a bank account or an investment plan) has multiple attributes. For instance, State Bank of India, the largest public sector bank in India, offers different kinds of savings accounts.\footnote{For the description of different savings accounts, see \url{https://www.sbi.co.in/web/personal-banking/accounts/saving-account}} Its regular savings account is different from its ``basic savings bank deposit small account" in a number of aspects, of which two are: (i) the regular account requires valid KYC (know your customer) documents, while the KYC requirement is relaxed for the small account, (ii) the balance is capped at Rupees 50,000 for the small account, while the regular account has no limit on maximum balance. For a poor customer living in a rural area, the transaction cost of obtaining  KYC documents is likely to be significant, while the cap is unlikely to matter. On the other hand, for a customer holding a well paying salaried job, obtaining KYC documents will not be a problem, but the cap on balance can be a major inconvenience. In both cases, in choosing between two accounts, a customer does not face a situation of marginal differences with respect to either of the two attributes (transaction cost of KYC and inconvenience of balance limit).          

A cursory look at popular online platforms such as Amazon and Expedia also shows that in many instances (although not always) a buyer does not face a situation of marginal differences. A product with a faster delivery time or an air ticket with a shorter duration of journey usually has a non-negligible difference in prices. As the rigidity of the noncompensatory property does not pose any problem in these examples, the lexicographic rule can be useful in these situations.  

This paper presents a characterization of lexicographic preferences in an individual decision making framework. The early literature can be traced back to Fishburn (1975), whose axiomatization of lexicographic preferences is closely connected with Arrow's (1951) impossibility theorem. For a lexicographic preference over a finite-dimensional product set, the most dominant factor is dictatorial in the sense that for two points that differ with respect to that factor, the preference order is solely determined by that specific factor; if two points are same with respect to the most dominant factor, the second most dominant factor is dictatorial and so on. In the words of Fishburn, there is a  ``hierarchy of dictators" for lexicographic preferences. Fishburn's proof begins by establishing the existence of a smallest decisive subset of the set of factors to show that this subset must contain a single element. A recent alternative proof by Mitra and Sen (2014) reconfirms the Arrow-Fishburn interconnections by determining an extremely pivotal factor along the same lines Geanakoplos (2005) identifies an extremely pivotal voter to prove Arrow's theorem.

More recently, Petri and Voorneveld (2016) propose another characterization of lexicographic preferences. They point out the unsatisfactory domain restriction in Fishburn's approach that is needed to apply Arrow's proof technique. Their characterization  is based on robustness of preference ordering between two alternatives for changes in a few rather than a large number of coordinates. However, it requires Fishburn's noncompensation axiom, which explicitly ``prohibits tradeoffs between factors".

We consider an individual decision problem over alternatives that are identified by a finite number $n$ of attributes, where the domain of each attribute is the set of all non-negative real numbers, so that the set of all alternatives is $\mathbb{R}^n_+.$ In this framework, we provide a different characterization of lexicographic preferences. Our analysis offers three key distinctions with the existing literature. First, in contrast with the previous works, continuity arguments play a prominent role in our analysis. Second, we follow the stepwise approach of looking at two attributes at a time, so our analysis mostly imposes structure on the induced preferences that are defined on $\mathbb{R}^2_+.$ Third, unlike Fishburn (1975) and Petri and Voorneveld (2016), we do not impose noncompensation on the preference; we consider an alternative condition (IMIA) that is weaker than noncompensation (see Section \ref{nonc}). 

\subsection{Use of continuity} The lexicographic preference is frequently used as a textbook example of a discontinuous preference. One main theoretical contribution of this paper is to show that by refining the notion of discontinuity, standard continuity techniques can be applied to understand lexicographic preferences.

Essentially,  discontinuity is an issue for a lexicographic preference when comparisons involve marginal differences. As discussed before, this is precisely where the noncompensatory property is problematic. If the available alternatives are such that when two alternatives differ with respect to an attribute, the difference is not marginal, then discontinuity is not an issue. Referring to the examples given before, if a buyer has chosen Galaxy A51 over Galaxy A71 following the lexicographic rule, she will not alter her choice if the price of Galaxy A71 drops slightly.  Similarly a customer who has chosen the small account over the regular account will not alter his choice if the maximum balance of the small account is lowered slightly. 

Formally, refining the notion of discontinuity amounts to weakening the notion of continuity to what we call mild continuity. To define mild continuity, call two alternatives totally different if they are different with respect to every attribute. A preference relation is mildly continuous if strict preference order between any two totally different alternatives is preserved around their small neighborhoods (Definition \ref{mcont}).\footnote{For the examples of smartphones and bank accounts, the alternatives differ with respect to each of the mentioned attributes, but they can be same in regard to some other attributes (e.g., battery life for phones, interest rate for bank accounts). Our analysis does not require mild continuity for the preference itself, but for the induced preferences where all but two attributes are fixed across alternatives (see Definition \ref{ind}, Axiom \ref{mild}), so it involves totally different alternatives with respect to a subset of attributes, as in the examples. It should be noted that mild continuity of induced preferences is neither necessary nor sufficient for a preference to be mildly continuous (see Section \ref{mildc}).} 

Discontinuity poses some problems for studying consumer behavior in discrete choice experiments (e.g., Gilbride and Allenby  2004, Campbell et al. 2006). In particular, for experiments on consumer behavior related to health care, such problems are sometimes addressed by simply deleting those responses in the data set that indicate lexicographic ordering or discontinuity in preference (e.g., McIntosh and Ryan 2002, Lancsar and Louviere 2006). Deleting a subset of responses from a data set may not be desirable.\footnote{See Bahrampour et al. (2020, p.389) for an overview of this issue for experiments in the context of health care.} Our approach can be useful to better understand the nature of consumer preferences for such deleted subsets (for instance, by seeing if mild continuity holds for those observations). 

Another related issue that arises in these experiments is the ambiguity between a {\it dominant} preference (that is, a preference in which the individual cares exclusively about only one attribute) and the lexicographic preference (e.g., Scott 2002,  Meenakshi et al. 2012). As strong monotonicity can be used to separate a dominant from a lexicographic preference,\footnote{For instance, on $\mathbb{R}^3_+,$ a dominant preference with respect to attribute $1$ is represented by utility function $u(x_1,x_2,x_3)=x_1$ (the individual cares exclusively about attribute $1$). As shown in Example \ref{ex4}, a dominant preference satisfies all of our axioms of lexicographic preference except strong monotonicity.} our results can be useful to clarify this ambiguity. A dominant preference is  continuous, so correctly specifying such a preference directly resolves the issue of discontinuity.

It should be mentioned that the extent to which our results are useful will depend on the specific context of an experiment. For instance, verifying strong monotonicity in an experiment seems relatively easy if an attribute is quantitative such as the price of a transport as in S{\ae}lensminder (2002) or the waiting time for a surgery as in McIntosh and Ryan (2002), but can be difficult if an attribute is qualitative in nature such as the color of maize|white, yellow or orange|as in the study of Meenakshi et al. (2012). Similarly, running additional tests on a subset of initially deleted observations maybe feasible sometimes, but too expensive in other situations.

\subsection{The results} 

We first consider the decision problem where alternatives have only two attributes. Together with mild continuity, the other central concept used in our analysis is the notion of an unhappy set, which an extension of lower contour sets. A set of alternatives is called an unhappy set if any alternative outside the set is preferred to all alternatives inside (Definition \ref{unhappyset}).\footnote{Unhappy sets were introduced in Mitra and Sen (2014).} We characterize lexicographic preferences with two attributes by imposing a certain structure on the unhappy sets. Consider two alternatives $a,b$ in an unhappy set. Suppose $b$ is superior in regard to attribute $1$ and $a$ is superior in regard to attribute $2.$ We say an unhappy set satisfies inclusion of marginally improved alternatives (IMIA) if it includes a third alternative $c$ that is same with $a$ in regard to attribute $2,$ but marginally improves upon $a$ in regard to attribute $1$ (Definition \ref{defn0}). We show that a complete and transitive preference relation on $\mathbb{R}^2_+$ is lexicographic if and only if it is strong monotone, mildly continuous and any closed unhappy set satisfies IMIA (Theorem \ref{cor3}). 

Extending to the general case of more than two attributes, the starting point is to compare alternatives for which all but two attributes have level zero.\footnote{This approach is closely related to choice rules based on ``elimination by aspects" proposed by Tversky (1972).} The zero level can be interpreted as the minimum or basic level of an attribute. Fix any subset of two attributes and consider all alternatives for which all attributes outside that subset have zero level. The original preference relation gives an induced preference between alternatives for which all but two fixed attributes have level zero. Such an induced preference is defined on $\mathbb{R}^2_+.$ By applying the characterization result for $\mathbb{R}^2_+,$ each of such induced preference is lexicographic if and only if each of them is strong monotone, mildly continuous and any closed unhappy set satisfies IMIA (Theorem \ref{tpair}). This gives a characterization of a ``pairwise lexicographic" preference (a preference for which any induced preference with two attributes fixing all other attributes at the minimum level zero is lexicographic). This result is of independent interest as such pairwise comparison can give some information about the relative importance of attributes for the decision maker. 

To get a lexicographic preference from the class of pairwise lexicographic preferences, we require nonreversibility under additional attributes (NRAA).  As before consider two alternatives for which all but two attributes have zero level. Suppose one of these alternatives is preferred to the other. Now add positive levels of one or more attributes to each of these alternatives keeping the levels of additional attributes same across the two. NRAA holds if the preference order between such new pairs of alternatives stays the same as before (Definition \ref{defn3}). Consider the example of smartphones in which any phone has three attributes: price, storage capacity and quality of camera. Consider two phones, each of which has zero level (that is, the minimum level) of storage capacity and suppose the first phone is preferred to the second. NRAA says that raising the storage capacity by the same amount to each of them without changing their prices or quality of camera does not alter the preference order.\footnote{Although NRAA has some resemblance with the independence axiom of Fishburn (1975), they are not same. See Section \ref{nrai} for examples that show that Fishburn's independence axiom is neither necessary nor sufficient for NRAA.}

In the general case of more than two attributes, we show that a complete and transitive preference relation is lexicographic if and only if (a) any induced preference between alternatives for which all but two same attributes have zero levels satisfies (i) strong monotonicity, (ii) mild continuity, (iii) IMIA for any closed unhappy set and (b) nonreversibility under additional attributes  holds (Theorem \ref{tone}). 

\subsection{Stepwise approach}

Bettman et al. (1998, p.189) point out two distinct ways of processing information in a decision problem involving products with multiple attributes:   

\begin{quote} 
``...information may be processed primarily by alternative, in which multiple attributes of a single option are processed before another option is considered, or by attribute, in which the values of several alternatives on a single attribute are examined before information on another attribute is considered."
\end{quote}
Our approach, where the decision maker compares alternatives by looking at two attributes at a time, more closely follows the second method above. This is consistent with the findings that consumers often screen products on the basis of one or two important attributes (e.g., Gilbride and Allenby 2004) and more generally, ``People...often look up at one or two relevant cues, avoid searching for conflicting evidence, and use noncompensatory strategies" (Goldstein and Gigerenzer 1999, p.82). Our characterization of lexicographic preferences proceeds in two steps, where the first step gives a structure on the induced preferences with two attributes, resulting in pairwise lexicographic preferences. Nonreversibilty, where additional attributes of same levels are added, gives lexicographic preference in the second step.

\subsection{IMIA and noncompensation}

The IMIA requirement (Axiom \ref{imia}) is a weaker condition than Fishburn's noncompensation axiom.\footnote{Consider any alternatives $x,y,w,z.$ The noncompensation condition says: suppose for every attribute $i,$ $x_i>y_i$ if and only if $w_i>z_i$ and $y_i>x_i$ if and only if $z_i>w_i;$ then the individual prefers $x$ to $y$ if and only if the individual prefers $w$ to $z.$} Specifically, we show that Fishburn's noncompensation implies noncompensation of induced preferences on $\mathbb{R}^2_+,$ which in turn implies the IMIA condition (Proposition \ref{nonc4}). 

Table 1 presents a comparison of these conditions for lexicographic and two related preferences: (i) pairwise lexicographic and (ii) the lexi-max preference, which follows a lexicographic order on magnitudes of attributes rather than their identities. The IMIA requirement, being the weakest of these three conditions, hold for all three preferences (see Section \ref{nonc} for details). 

\newpage
{\bf Table 1 Examples of preferences satisfying different conditions}
{\small{
\begin{center}
\begin{tabular}{|c|c|c|c|}
\cline{1-4} {} & {Lexicographic} & {Pairwise lexicographic,} & {Lexi-max} \\
{} & {preference} & {but not lexicographic} & {preference} \\
{} & {} & {(Example \ref{ex2})} & {(Example \ref{ex50})} \\

\cline{1-4} {Fishburn's} & {$\checkmark$} & {} & {} \\

{noncompensation axiom} & {} & {} & {} \\
\cline{1-4} {Noncompensation for} & {$\checkmark$} & {$\checkmark$} & {} \\
{induced preferences} & {} & {} & {} \\
{on $\mathbb{R}^2_+$ (Axiom \ref{conb})} & {} & {} & {} \\
\cline{1-4}
{IMIA requirement} & {$\checkmark$} &{$\checkmark$}& {$\checkmark$} \\
{(Axiom \ref{imia})} & {} & {} & {} \\
\cline{1-4}
\end{tabular}
\end{center}}}

While we theoretically establish that IMIA is a less demanding requirement, it can be argued that this condition, which involves extensions of lower contour sets, may not be straightforward to use for applications. We address this issue by proposing an alternative characterization where the IMIA (Axiom \ref{imia}) is replaced by noncompensation of induced preferences on $\mathbb{R}^2_+$ (Axiom \ref{conb}), which is a stronger requirement, but still less stringent than Fishburn's noncompensation axiom. We show that our characterization of pairwise lexicographic and lexicographic preferences go through with  Axiom \ref{conb} (Corollary \ref{talt}). Since Axiom \ref{conb} is relatively simple to explain and verify, this alternative characterization may be more suitable to use for applications such as designing experiments.

The paper is organized as follows. We present the analytical framework in Section \ref{frame} where the key concepts are introduced. Section \ref{lextwo} looks at lexicographic preferences with two attributes and presents the characterization result (Theorem \ref{cor3}). Section \ref{lexmore} considers more than two  attributes, where we present the result on pairwise lexicographic preferences (Theorem \ref{tpair}) and the main result (Theorem \ref{tone}). In Section \ref{example} we discuss the implications of the axioms. Most proofs are presented in the Appendix.
	
\section{The analytical framework}
\label{frame}

\hyperref[sec:frame]{}

Consider an individual who has a preference relation $\succsim$ on a set of alternatives $X.$ Each alternative is characterized by $n$ attributes. Let $N=\{1,\ldots,n\}$ be the set of attributes. The domain of any attribute $i\in N$ is $\mathbb{R}_+.$ Therefore an alternative is given by a vector $x=(x_1,\ldots,x_n)\in \mathbb{R}^n_+$ and the set of all alternatives is $X=\mathbb{R}^n_+.$

For any non empty $S\subseteq N,$ denote $X_S=\mathbb{R}^{|S|}_+$. For any $x\in X$ and $S\subseteq N,$ we write $x=(x^S,x^{N\setminus S})$ where $x^S\in X_S$ and $x^{N\setminus S}\in X_{N\setminus S}$ (note that $x=x^N$). If $x_i=0$ for all $i\in S,$ we write $x^S=0^S.$

In the special case when $S$ is the singleton set $\{i\},$ it will be convenient to use the simpler notation $x^S=x_i,$ $x^{N\sm S}=x_{-i},$ $X_S=X_i$ and $X_{N\setminus S}=X_{-i}.$ We write $x=(x_i,x_{-i})$ where $x_i\in X_i$ and $x_{-i}\in X_{-i}.$

The distance between $x,y\in X,$ denoted by $d(x,y),$ is given by the Euclidean metric. For $x^S,y^S\in X_S,$ $d(x^S,y^S)$ is the same metric $d$ restricted to $X_S.$ A neighborhood of $x^S$ is a set $B_{\ve}(x^S)$ consisting of all $y^S\in S$ such that $d(x^S,y^S)<\ve$ for some $\ve>0.$

The individual's preference on $X$ is defined using the binary relation $\succsim$ where  ``$x\succsim y$" stands for ``the individual prefers $x$ to $y$". The strict preference ``$x\succ y$" stands for ``the individual strictly prefers $x$ to $y$" and is defined as $x\succ y\Leftrightarrow$ $[x\succsim y]$ and $[\mbox{not }y\succsim x].$ The indifference relation ``$x\sim y$" stands for ``the individual is indifferent between $x$ and $y$" and is defined as  $x\sim y\Leftrightarrow [x\succsim y]$ and $[y\succsim x].$

For any $x\in X$, the lower contour set of $x$ under $\succsim$ is $L(x)=\{y\in X\mid x\succsim y\}.$ The strict lower contour set of $x$ is $\ul{L}(x)=\{y\in X\mid x\succ y\}$ and the indifference set of $x$ is $I(x)=\{y\in X\mid x\sim y\}$.

A preference relation $\succsim$ on $X$ is {\it complete} if for any $x,y\in X,$ either $x\succsim y$ or $y\succsim x.$ It is {\it transitive} if for any $x,y,z\in X,$ whenever $x\succsim y$ and $y\succsim z,$ we have $x\succsim z.$ Throughout we consider preference relations on $X$ that are complete and transitive.

Let $x,y\in X.$ If $x_i>y_i$ for all $i\in N$, we write $x>y.$ If $x_i\geq y_i$ for all $i\in N,$  we write $x\geq y.$ A preference relation $\succsim$ on $X$ is {\it monotone} if for any $x,y\in X$ with $x>y$, we have $x\succ y$. It is {\it strong monotone} if for any $x,y\in X$ with $x\geq y$ and $x\neq y$, we have $x\succ y$.

\begin{defn}\label{lex}
	Let $N=\{1,\ldots,n\}$ be the set of attributes. A preference relation $\succsim$ on $X= \mathbb{R}^n_+$ is {\it lexicographic} if $x\sim x$ for all $x\in X$ and the set of attributes can be written as $N=\{i_1,\ldots,i_n\}$ such that for any $x,y\in X,$ $x\succ y$ if and only if either
[$x_{i_1}>y_{i_1}$] or [$x_{i_1}=y_{i_1},$ $x_{i_2}>y_{i_2}$] or \ldots or [$x_{i_1}=y_{i_1}$, \ldots, $x_{i_{n-1}}=y_{i_{n-1}},$ $x_{i_n}>y_{i_n}$]. For this preference $i_1$ is the most important attribute, $i_2$ the next most important attribute and so on and the preference is denoted by 
$i_1\succ^L\ldots\succ^Li_n.$  
\end{defn}

\subsection{Some useful concepts}

To characterize lexicographic preferences, it will be useful to introduce (i) mild continuity of a preference relation, which is a weaker version of continuity, (ii) unhappy sets, which are related to lower contour sets and (iii) induced preferences.

\subsubsection{Mild continuity}

\begin{defn} \label{td}
	\rm For $x,y\in X,$ we say $x$ and $y$ are {\it totally different}, denoted by $x \td y,$ if $x_i\neq y_i$ for all $i\in N.$
\end{defn}Note that $x \td y$ if and only if $y \td x.$ For $x^S,y^S\in X_S,$ we define $x^S\td y^S$ similarly. A preference relation is mildly continuous if strict preference order between any two totally different points is preserved around their small neighborhoods.

\begin{defn} \label{mcont}
	\rm A preference relation $\succsim$  is {\it mildly continuous} on $X$ if for any $x,y\in X$ with $x \td y$ and $x\succ y$, there exists $\ve>0$ such that if $\wt{x}\in B_{\ve}(x)$ and $\wt{y}\in B_{\ve}(y),$ then $\wt{x}\succ \wt{y}.$
\end{defn}We recall that a preference relation $\succsim$ is continuous on $X$  if for any $x,y\in X$ with $x\succ y$, there exists $\ve>0$ such that if $\wt{x}\in B_{\ve}(x)$ and $\wt{y}\in B_{\ve}(y),$ then $\wt{x}\succ \wt{y}.$ Thus for a continuous preference, strict preference order between any two points, totally different or otherwise, is preserved around their small neighborhoods.

\subsubsection{Unhappy sets}

\begin{defn} \label{unhappyset}
	\rm A set $A\subseteq X$ is an {\it unhappy set for a preference relation $\succsim$} on $X$ if for any $y\in X\setminus A$, $y\succ x$ for every $x\in A$.
\end{defn}Observe that lower contour and strict lower contour sets are unhappy sets. So are the sets $X$ and $\emp.$

Let $A$ be a subset of a metric space $X.$ A point $x\in X$ is a boundary point of $A$ if every neighborhood of $x$ contains at least one point in $A$ and at least one point in $X\sm A.$ The set of all boundary points of $A$ is called the boundary of $A$ and denoted by $\partial{A}.$ To characterize unhappy sets we recall the following result. For the proof, see, e.g., Mendelson (1990: Theorem 4.23, Chapter 3 and Definition 2.1, Chapter 4).

\begin{res} \label{res1}{\it If $A$ is a non empty proper subset of a connected space, then {\rm(i)} ${\partial A}\neq \emp$ and {\rm (ii)} the set $A$ cannot be both open and closed.}
\end{res}

Since $\mathbb{R}^n_+$ is a connected set, we can use this result for subsets of $X=\mathbb{R}^n_+.$

\begin{prop}\label{propone}
{\it Consider a complete and transitive preference relation $\succsim$ on $X=\mathbb{R}^n_+.$}
\begin{enumerate}[\rm(i)]
\item {\it If $A,B$ are unhappy sets, then either $A\subseteq B$ or $B\subseteq A.$}
\item {\it Let $A$ be a non empty proper subset of $X.$ If $A$ is an unhappy set and $\succsim$ is mildly continuous, then for any $x\in {\partial A},$ the following hold for any $y\td x.$} \begin{enumerate}[\rm(a)]
\item {\it If $y\in {\partial A},$ then $x\sim y.$}
\item {\it If $y\in A,$ then $x\succsim y.$}
\item {\it If $x\succ y,$ then $y\in A.$}
\end{enumerate}
\end{enumerate}
\end{prop}\noi {\bf Proof} See the Appendix. \qed

\vspace{0.1cm} For a continuous preference relation, the last part of the proposition also holds for $y$ that are not totally different from $x.$ In that case unhappy sets can be more precisely characterized.

\begin{cor}\label{cor1}{\it Consider a complete, transitive and continuous preference relation $\succsim$ on $X=\mathbb{R}^n_+.$ Let $A$ be an unhappy set which is a non empty proper subset of $X.$ Then the following hold for  any $x\in {\partial A}.$}
\begin{enumerate}[(i)]
\item {\it $x\sim y$ for any $y\in \partial{A}.$}
\item {\it $\ul{L}(x)\subseteq A \subseteq L(x).$}
\item {\it The set $A$ must be either closed or open, but not both. If $A$ is closed, then $A=L(x)$ and if $A$ is open, then $A=\ul{L}(x).$}
\end{enumerate}\end{cor}\noi {\bf Proof} See the Appendix. \qed

\vspace{0.1cm} Thus for a continuous preference a set is an unhappy set if and only if it is a lower contour or a strict lower contour set. A lexicographic preference is not continuous, although it is mildly continuous. Any lower contour set of a lexicographic preference is neither open nor closed.

\subsubsection{Induced preferences}

Fix a subset of attributes and consider all points for which attributes in that subset have zero level. The original preference relation gives a preference order between alternatives for which a fixed subset of attributes have level zero. This is formalized by the notion of an induced preference.

\begin{defn} \label{ind}
	\rm Let $S$ be a non empty subset of $N.$ For a preference relation $\succsim$ on $X,$  the {\it induced preference} $\succsim_S$ on $X_S=\mathbb{R}^{|S|}_+$ is defined as follows: for $y^S,z^S\in X_S,$ $y^S\succsim_S z^S$ if and only if $(y^S,0^{N\setminus S})\succsim (z^S,0^{N\setminus S}).$
\end{defn}Thus $\succsim_S$ is a preference relation over all alternatives for which the attributes in the set $N\sm S$ have zero level. Note that the induced preference $\succsim_N$ coincides with $\succsim.$ 

\begin{rem}\label{remlex} For a lexicographic preference, any induced preference is also lexicographic.
If $\succsim$ is the lexicographic preference $1\succ^L\ldots\succ^Ln$ and $S=\{i_1,\ldots,i_s\}\subseteq N$ where $i_1<\ldots<i_s,$ then $\succsim_S$ is the lexicographic preference $i_1\succ^L\ldots\succ^Li_s.$ 
\end{rem} 

We can define unhappy sets for induced preferences. We say a set $A\subseteq X_S$ is an unhappy set for the induced preference $\succsim_S$ if for any $b\in X_S\setminus A$, $b\succ_S a$ for every $a\in A.$ For $x\in X$ and a preference relation $\succsim$ on $X,$ denote by $\ol{L}(x)$ the closure of the lower contour set $L(x),$ that is, $\ol{L}(x):=L(x)\cup \partial{L(x)}.$ The next proposition shows that for a complete, transitive and strong monotone preference, mild continuity of induced preferences ensures that the closure of a lower contour set is an unhappy set.

\begin{prop}\label{proptwo}
{\it Consider a complete, transitive and strong monotone preference relation $\succsim$ on $X=\mathbb{R}^n_+.$ Suppose for any $S\subseteq N,$ the induced preference $\succsim_S$ is mildly continuous on $X_S.$ Then for every $x\in X,$ the set $\ol{L}(x)$ is an unhappy set for $\succsim.$}
\end{prop}\noi {\bf Proof} See the Appendix. \qed

\vspace{0.1cm} A strong monotone preference relation on $\mathbb{R}_+$ is continuous; so it is mildly continuous. Suppose $\succsim$ is strong monotone and let $S\subseteq N$ be a singleton set. Then the induced preference $\succsim_S,$ which is defined on $\mathbb{R}_+,$ is strong monotone and therefore mildly continuous. When $X=\mathbb{R}^2_+$ in Proposition \ref{proptwo}, then any non empty proper subset of $X$ is a singleton set and mild continuity already holds for the corresponding induced preference. This gives the following result.

\begin{cor}\label{cor2}{\it For a complete, transitive, strong monotone and mildly continuous preference relation on $\mathbb{R}^2_+,$ the closure of any lower contour set is an unhappy set.}\end{cor}

\section{Lexicographic preferences with two attributes}
\label{lextwo}

\hyperref[sec:lextwo]{}

We begin by presenting a characterization of lexicographic preferences with two attributes. This result is of independent interest. Moreover, it will be also useful for characterizing lexicographic preferences with more than two attributes.

\subsection{Inclusion of marginally improved alternatives}
\label{imo}

\hyperref[sec:imo]{}

Consider the individual decision problem where each alternative has only two attributes. In this case the  set of attributes is $N=\{1,2\}$ and the set of alternatives is $X=\mathbb{R}^2_+.$ 

\begin{defn} \label{defn0}
	\rm For a monotone preference relation $\succsim$ on $X=\mathbb{R}^2_+,$ an unhappy set $A$ satisfies {\it inclusion of marginally improved alternatives} (IMIA) if the following hold for any $x,y\in A$ with $x\td y$:

\noi (i) if $y>x,$ then $\exists$ $y>\wt{x}>x$ such that $\wt{x}\in A.$ 

\noi (ii) if $y_2>x_2$ and $x_1>y_1,$ then $\exists$ $x_2<\wt{x}_2<y_2$ such that $(x_1,\wt{x}_2)\in A.$

\noi (iii) if $y_1>x_1$ and $x_2>y_2,$ then $\exists$ $x_1<\wt{x}_1<y_1$ such that $(\wt{x}_1,x_2)\in A.$ 

\end{defn}

Suppose $x,y$ are in an unhappy set $A.$ Since the preference relation is monotone, if $y>x,$ then $\wt{x}\in A$ {\it for any} $y>\wt{x}.$ This shows Definition \ref{defn0} does not impose any additional requirement when $y>x.$ 

\begin{figure}[!h]\centering\includegraphics[width=5.5in]{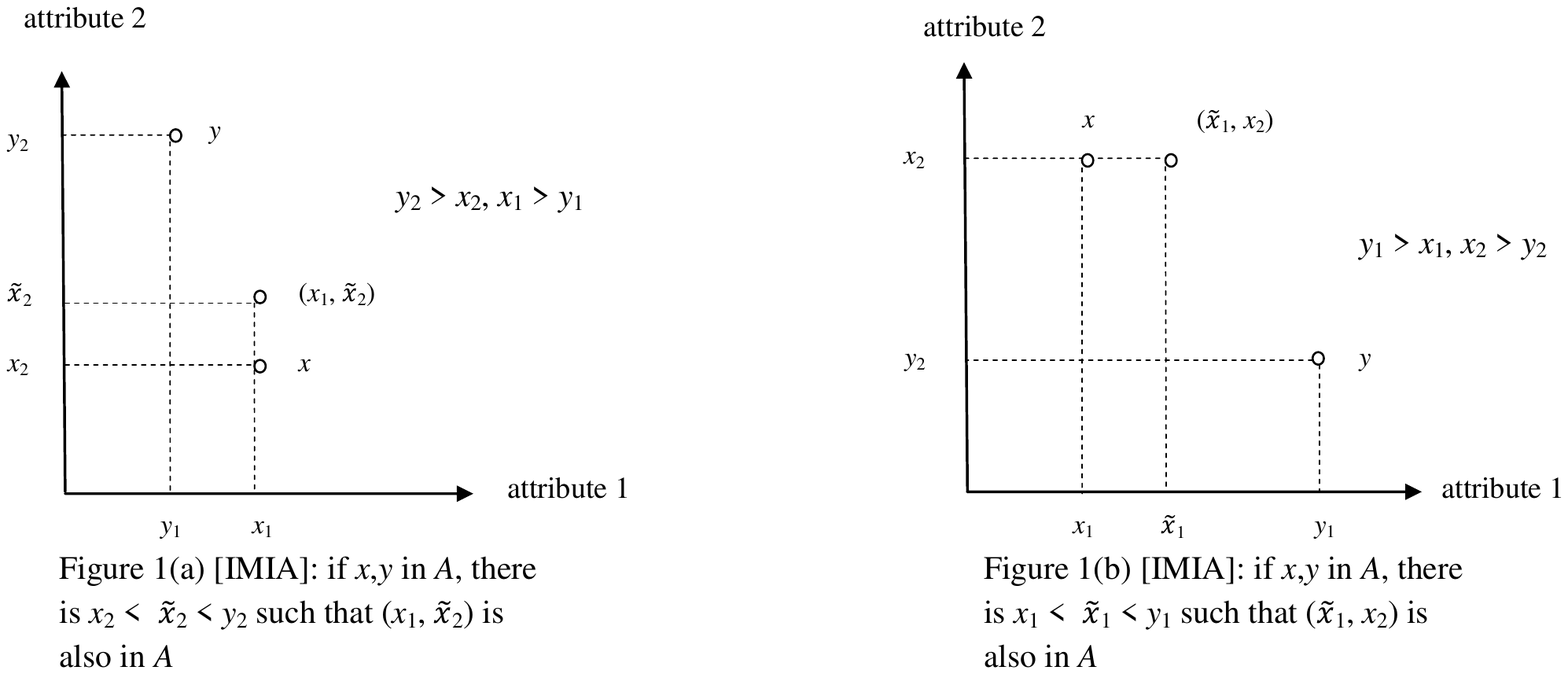}\end{figure}

If $y_2>x_2$ and $x_1>y_1,$ then Definition \ref{defn0} requires there exists $x_2<\wt{x}_2<y_2$ such that $(x_1,\wt{x}_2)\in A$ (see Figure 1(a)). If such a $\wt{x}_2$ exists, then by the monotonicity of the preference relation, $(x_1,z_2)\in A$ for all $x_2<z_2<\wt{x}_2.$ This means if $x,y$ are in an unhappy set $A$ where $x$ is superior in regard to the first and inferior in regard to the second attribute, then alternatives that are marginal improvements over $x$ in regard to the second  attribute (specifically, alternatives $(x_1,z_2)$ with $x_2<z_2<\wt{x}_2$) are also in $A.$ 

Similarly if $y_1>x_1$ and $x_2>y_2,$ then Definition \ref{defn0} requires there exists $x_1<\wt{x}_1<y_1$ such that $(\wt{x}_1,x_2)\in A$ (see Figure 1(b)). If such a $\wt{x}_1$ exists, then by monotonicity, $(z_1,x_2)\in A$ for all $x_1<z_1<\wt{x}_1.$ This means if $x,y$ are in an unhappy set $A$ where $x$ is superior in regard to the second and inferior in regard to the first attribute, then alternatives that are marginal improvements over $x$ in regard to the first attribute are also in $A.$ 

\begin{rem}\label{remi} If $x=(x_1,x_2)$ is an interior point of $A,$ then there is a neighborhood $B_\ve(x)\subset A.$ In this case there always exist $\wt{x}_1>x_1,\wt{x}_2>x_2$ such that $(x_1,\wt{x}_2),(\wt{x}_1,x_2)\in B_\ve(x).$ Thus IMIA always holds for open unhappy sets such as strict lower contour sets. IMIA imposes additional structure on the preference only for unhappy sets that are not open. For example, for the continuous preference on $X=\mathbb{R}^2_+$ represented by the Cobb-Douglas utility function\footnote{A function $u: X\rightarrow \mathbb{R}$ is a {\it utility function representing} $\succsim$ if for all $x,y\in X,$ $x\succsim y$ if and only if $u(x)\geq u(y).$ Debreu's theorem ensures that any continuous preference relation is represented by some continuous utility function (see Debreu 1954 and also Lecture 2 of Rubinstein 2012).} 
 $u(x_1,x_2)=x^{\alpha}_1x^{1-\alpha}_2$ (where $0<\alpha<1$), IMIA does not hold for lower contour sets. For a lexicographic preference on $\mathbb{R}^2_+,$ IMIA does not hold for lower contour sets (which are neither open nor closed). However, any closed unhappy set of a lexicographic preference on $\mathbb{R}^2_+$  satisfies this property.
\end{rem} 

\begin{lem}\label{leximia}
{\it For a lexicographic preference on $X=\mathbb{R}^2_+,$ any closed unhappy set satisfies IMIA.}
\end{lem}\noi {\bf Proof} See the Appendix. \qed

\vspace{0.1cm} For a strong monotone and mildly continuous preference $\succsim$ on $X=\mathbb{R}^2_+,$ when a closed unhappy set satisfies IMIA, it has a certain structure as stated in Lemma \ref{bound}. Let $\mathcal{A}_{\succsim}$ be a family of subsets of $X$ defined as follows:
\be\label{a}\mathcal{A}_{\succsim}:=\{A|A\subset X;\exists z\in A\mbox{ with }z>(0,0);A\mbox{ is a closed unhappy set satisfying IMIA}\}\ee

\begin{lem}\label{bound}
{\it Consider a complete, transitive, strong monotone and mildly continuous preference relation $\succsim$ on $X=\mathbb{R}^2_+.$ Let $A\in \mathcal{A}_{\succsim}.$ Then $\exists$ a positive number $\alpha^A$ and an attribute $i^*\in \{1,2\}$ such that $A=\{x\in X|0\leq x_{i^*}\leq \alpha^A\}.$ Moreover the attribute $i^*$ is the same for all sets in $\mathcal{A}_{\succsim}.$}
\end{lem}\noi {\bf Proof} See the Appendix. \qed

\vspace{0.1cm} Applying Lemma \ref{bound}  and Corollary \ref{cor2}, we can characterize lexicographic preferences with two attributes. 

\begin{thm}\label{cor3}
{\it A complete and transitive preference relation $\succsim$ on $\mathbb{R}^2_+$ is lexicographic if and only if it is strong monotone, mildly continuous and any closed unhappy set of $\succsim$ satisfies IMIA.}
\end{thm}\noi {\bf Proof} See the Appendix. \qed

\section{Lexicographic preference with more than two attributes}
\label{lexmore}

\hyperref[sec:lexmore]{}

Our approach to characterizing lexicographic preferences with more than two attributes is through conditions on induced preferences $\succsim_S$ with $|S|=2$ that are defined on $\mathbb{R}^2_+,$ where all but two attributes are held fixed at the minimum level zero. This can be viewed as a stepwise process where the individual decision maker considers two attributes at a time. 

\subsection{Pairwise lexicographic preferences}

For the decision problem with two attributes, we know by Theorem \ref{cor3} that strong monotonicity, mild continuity and IMIA for any closed unhappy set ensure the preference is lexicographic. An immediate consequence of this result is if these properties are imposed on the induced preference $\succsim_S$ for every $S\subseteq N$ with $|S|=2,$ then each of such induced preferences will be lexicographic, which gives a ``pairwise lexicographic" preference, as defined below. 

\begin{defn}\label{dpair}
A preference relation $\succsim$ on $\mathbb{R}^n_+$ is {\it pairwise lexicographic} if  for every $S\subseteq N$ with $|S|=2,$ the induced preference  $\succsim_S$ is a lexicographic preference on $\mathbb{R}^2_+.$
\end{defn}
\begin{rem} \label{pair0}
If $\succsim$ is pairwise lexicographic, then for any $S\subseteq N$ with $|S|=2,$ there are $i,j\in N$ such that $S=\{i,j\}$ and $i$ is the most important attribute of $\succsim_S,$ that is, 
$x^S\succ_S y^S$ if and only if either $[x_i>y_i]$ or $[x_i=y_i,x_j>y_j]$ which means 
\be\label{pair1}(x^S,0^{N/S})\succ (y^S,0^{N/S})\mbox{ if and only if either }[x_i>y_i]\mbox{ or }[x_i=y_i,x_j>y_j]\ee 
If (\ref{pair1}) holds for $S=\{i,j\},$ we write $i\succ^* j.$ Note that if $\succsim$ is  pairwise lexicographic, then for any $i,j\in N,$ either $i\succ^* j$ or $j\succ^* i.$  
\end{rem}

\begin{ax} \label{mon}
\rm For any $S\subseteq N$ with $|S|=2,$ the induced preference $\succsim_S$ is strong monotone.
\end{ax}

\begin{ax} \label{mild} \rm For any $S\subseteq N$ with $|S|=2,$ the induced preference $\succsim_S$ is mildly continuous on $X_S=\mathbb{R}^2_+.$
\end{ax}

\begin{ax} \label{imia}\rm
For any $S\subseteq N$ with $|S|=2,$ any closed subset of $X_S=\mathbb{R}^2_+$ that is an unhappy set for the preference $\succsim_S,$ satisfies IMIA.	
\end{ax}

\begin{thm} \label{tpair}
{\it Consider any complete and transitive preference relation $\succsim$ on $X=\mathbb{R}^n_+$ where $n\geq 3.$ The following statements are equivalent.}
\begin{enumerate}
\item[\rm (P1)] {\it The preference relation $\succsim$ satisfies Axiom $\ref{mon},$ Axiom $\ref{mild}$ and Axiom $\ref{imia}.$}
\item[\rm (P2)] {\it The preference relation $\succsim$ is pairwise lexicographic.}
\end{enumerate}
 \end{thm}\noi{\bf Proof} Since the preference relation $\succsim$ is complete and transitive, so is the induced preference $\succsim_S$ for any $S\subseteq N$ with $|S|=2.$ Then the result is immediate from Theorem \ref{cor3}. \qed 

\vspace{0.1cm} Clearly a lexicographic preference is pairwise lexicographic, but the converse is not true (see Example \ref{ex2}). We need some additional structure to get a lexicographic preference  from the class of pairwise lexicographic preferences.

\subsection{Nonreversibility under additional attributes}

As before consider two alternatives for which all but two attributes have zero level and suppose the individual strictly prefers one of these to the other. Now suppose we raise the levels of one or more attributes that initially had zero levels, still keeping the levels of these attributes same across the two. Nonreversibility under additional attributes holds if the preference order between such new pairs of alternatives stays the same as before.

\begin{defn} \label{defn3}
	\rm A preference relation $\succsim$ on $X$ satisfies {\it nonreversibility under additional attributes} (NRAA) if for any $S\subseteq N$ with $|S|=2,$ the following hold: if $(x^S,0^{N/S})\succ (y^S,0^{N/S}),$ then $(x^S,z^{N/S})\succ (y^S,z^{N/S})$ for any $z^{N/S}\in X_{N\sm S}.$
\end{defn}
\begin{ax} \label{irev}
The preference relation $\succsim$ satisfies NRAA.	
\end{ax}

\begin{lem}\label{lex1}
{\it Consider a complete and transitive preference relation $\succsim$ on $X=\mathbb{R}^n_+$ that is pairwise lexicographic and satisfies Axiom $\ref{irev}.$ Then for any $i,j\in N,$ either $i\succ^* j$ or $j\succ^*i$ and the following hold.}

\begin{enumerate}[\rm(i)]

\item {\it Let $S=\{i,j\}$ and $i\succ^*j.$ If either $[x_i>y_i]$ or $[x_i=y_i$ and $x_j>y_j],$ then $(x^S,z^{N\sm S})\succ (y^S,z^{N\sm S})$for any $z^{N\sm S}\in X_{N\sm S}.$} 

\item {\it For $i,j,k\in N,$ if $i\succ^* j$ and $j\succ^* k,$ then $i\succ^*k.$}

\item {\it Based on $\succ^*,$ the attributes of $N$ can be ordered. That is, we can write $N=\{i_1,\ldots,i_n\}$ such that $i_1\succ^*\ldots\succ^*i_n.$}

\end{enumerate}\end{lem}\noi {\bf Proof} Part (i) follows from (\ref{pair1}) by applying Axiom \ref{irev}. See the Appendix for the proofs of parts (ii)-(iii). \qed

\subsection{The main result}

\begin{thm} \label{tone}
{\it Consider any complete and transitive preference relation $\succsim$ on $X=\mathbb{R}^n_+$ where $n\geq 3.$ The following statements are equivalent.}
\begin{enumerate}
\item[(L1)] {\it The preference relation $\succsim$ satisfies Axiom $\ref{mon},$  Axiom $\ref{mild},$ Axiom $\ref{imia}$ and Axiom $\ref{irev}.$}
\item[(L2)] {\it The preference relation $\succsim$ is lexicographic.}
\end{enumerate}
 \end{thm}\noi {\bf Proof} (L2)$\Rightarrow$(L1): Consider a lexicographic preference $\succsim$ on $\mathbb{R}^n_+$ for which without loss of generality (w.l.o.g.), $1$ is the most important attribute, $2$ is the second most important attribute and so on. Consider any $S\subseteq N$ with $|S|=2.$ If $(x^S,0^{N\sm S})\succ (y^S,0^{N\sm S}),$ then there are attributes $i<j$ such that $S=\{i,j\}$ and either (i) $[x_i>y_i]$ or (ii) $[x_i=y_i,x_j>y_j].$ This implies $(x^S,z^{N\sm S})\succ (y^S,z^{N\sm S})$ for any $z^{N\sm S},$ so Axiom \ref{irev} holds. As a lexicographic preference is also pairwise lexicographic, by Theorem \ref{tpair}, Axioms \ref{mon}-\ref{imia} also hold.

(L1)$\Rightarrow$(L2): Since Axioms \ref{mon}-\ref{imia} hold, by Theorem \ref{tpair}, the preference is pairwise lexicographic. As Axiom \ref{irev} holds, by Lemma \ref{lex1}(iii) we can write $N=\{i_1,\ldots,i_n\}$ such that $i_1\succ^*\ldots\succ^*i_n.$ W.l.o.g., let $i_1=1,\ldots,i_n=n.$ We prove that  $\succsim$ is the lexicographic preference $1\succ^L\ldots\succ^Ln$ by showing that for any $m\in \{1,\ldots,n\},$ if $x,y$ are such that $x_i=y_i$ for all $i<m$ and $x_m>y_m,$ then $x\succ y.$ 

First let $m=n$ and suppose $x,y$ are such that $x_i=y_i$ for all $i<n$ and $x_n>y_n.$ Since $n-1\succ^*n,$ in this case by Lemma \ref{lex1}(i), we have $x\succ y.$  

Next consider any $m\in\{1,\ldots,n-1\}.$ Let $S=\{i\in N|i<m\}$ and $T=\{i\in N|i>m\}.$ Let $x,y\in X$ be such that $x^S=y^S$ and $x_m>y_m.$ We can find $x^0_m,\ldots,x^{n-m}_m\in \mathbb{R}_+$ such that $x_m=x^0_m>x^1_m>\ldots>x^{n-m}_m=y_m.$ Construct $z[k]=(z[k]_1,\ldots,z[k]_n)\in X$ recursively as follows: $z[0]=x$ and for $k=1,\ldots,n-m,$ $z[k]$ is such that
$$z[k]_m=x^k_m,z[k]_{m+k}=y_{m+k},{z[k]}^S={z[0]}^S=x^S\mbox{ and }{z[k]}^{T\sm \{m+k\}}={z[k-1]}^{T\sm \{m+k\}}$$
Thus
$$z[1]_i=x_i\mbox{ for }i<m,z[1]_m=x^1_m,z[1]_{m+1}=y_{m+1},z[1]_i=x_i\mbox{ for }i>m+1$$
$$z[2]_i=x_i\mbox{ for }i<m,z[2]_m=x^2_m,z[2]_i=y_i\mbox{ for }i=m+1,m+2,z[2]_i=x_i\mbox{ for }i>m+2$$
and in general
$$z[k]_i=x_i\mbox{ for }i<m,z[k]_m=x^k_m,z[k]_i=y_i\mbox{ for }i=m+1,\ldots,m+k,z[k]_i=x_i\mbox{ for }i>m+k$$
Observe that $z[k]$ and $z[k-1]$ have same levels for all but two attributes ($m$ and $m+k$). Since $m\succ^*{m+k}$ and $z[k-1]_m=x^{k-1}_m>z[k]_m=x^k_m,$ by Lemma \ref{lex1}(i) we have $z[k-1]\succ z[k].$ Noting that $z[n-m]=y,$ we conclude that $x=z[0]\succ z[1]\succ \ldots\succ z[n-m]=y.$ Hence $x\succ y.$

Thus for any $x,y$ with $x_i=y_i$ for all $i<m$ and $x_m>y_m,$ we have $x\succ y.$ Applying this result for $m=1,\ldots,n,$ proves (L2). \qed

\section{Discussion on the axioms}
\label{example}

\hyperref[sec:example]{}

In conclusion, we discuss the implications of our axioms in reference to the axioms of Fishburn (1975) and also look at their robustness.

\subsection{Fishburn's noncompensatory axiom and IMIA}

\label{nonc}

\hyperref[sec:nonc]{}

For a strong montone preference relation $\succsim$ on $X=\mathbb{R}^n_+,$ Fishburn's noncompensatory axiom (see Axiom 3 of Fishburn 1975) can be stated as: if ($x_i>y_i$ if and only if $z_i>w_i$) and ($y_i>x_i$ if and only if $w_i>z_i$), then ($x\succ y$ if and only if $z\succ w$) and ($y\succ x$ if and only if $w\succ z$) for all $x,y,z,w\in X.$ For a concise presentation of this condition, for $x,y\in X,$ let
\be\label{nonc1}M(x,y):=\{i\in N|x_i>y_i\}\mbox{ and }E(x,y):=\{i\in N|x_i=y_i\}\ee Note that if $M(x,y)=M(z,w)$ and $M(y,x)=M(w,z),$ then $E(x,y)=E(z,w).$ 

\begin{defn} \label{nonc2}
	\rm A strong monotone preference relation $\succsim$ on $X=\mathbb{R}^n_+$ is {\it noncompensatory} if it satisfies Fishburn's noncompensatory axiom, that is, for all $x,y,z,w\in X$: 
$$\mbox{ if }M(x,y)=M(z,w)\mbox{ and }M(y,x)=M(w,z),\mbox{ then }(x\succ y\Leftrightarrow z\succ w)\mbox{ and }(y\succ x\Leftrightarrow w\succ z).$$
\end{defn}

\begin{rem} \label{nonc3}
	\rm For any non empty $S\subset N,$ if $x=(x^S,0^{N/S})$ and $y=(y^S,0^{N/S}),$ then $M(x,y)=M(x^S,y^S)$ and $M(y,x)=M(y^S,x^S).$ This shows if a preference relation $\succsim$ is noncompensatory, then so is the induced preference $\succsim_S$. However, the converse is not true, as shown in the following example. 
\end{rem}

\begin{ex}\label{ex2} Consider a preference relation $\succsim$ on $X=\mathbb{R}^3_+$ that is reflexive ($x\sim x$ for all $x\in X$) and for which following hold for any $x_i,y_i\in \mathbb{R}_+$:
(a) $(x_1,x_2,x_3)\succ (y_1,y_2,y_3)$ if $x_1>y_1,$ (b) $(0,x_2,x_3)\succ (0,y_2,y_3)$ if $x_2>y_2,$ (c) $(0,x_2,x_3)\succ (0,x_2,y_3)$ if $x_3>y_3,$ (d) $(x_1,x_2,x_3)\succ (x_1,y_2,y_3)$ if $x_1>0$ and $x_2+x_3>y_2+y_3$ and (e) $(x_1,x_2,x_3)\sim (x_1,y_2,y_3)$ if $x_1>0$ and $x_2+x_3=y_2+y_3.$ This preference is pairwise lexicographic, but not lexicographic. 

For any $S\subseteq N$ with $|S|=2,$ the induced preference $\succsim_S$ is lexicographic, so $\succsim_S$ is noncompensatory. 
Take $x=(0,6,1),$ $y=(0,4,8),$ $z=(1,5,4),$ $w=(1,4,7).$ Note that $M(x,y)=M(z,w)=\{2\}$ and $M(y,x)=M(w,z)=\{3\}.$ We have $x\succ y$ (by (b)), but $w\succ z$ (by (d)), so $\succsim$ is not noncompensatory.

\end{ex}

The next proposition shows that Fishburn's noncompensatory axiom implies Axiom \ref{imia}.	

\begin{prop}\label{nonc4}{\it Consider the following statements for a complete, transtitive and strong montone preference relation $\succsim$ on $X=\mathbb{R}^n_+.$}

\begin{enumerate}[(A)]

\item {\it $\succsim$ is noncompensatory.}

\item {\it For any $S\subseteq N$ with $|S|=2,$ the induced preference $\succsim_S$ is noncompensatory.} 

\item {\it Axiom $\ref{imia}$ holds, that is, for any $S\subseteq N$ with $|S|=2,$ any closed subset of $X_S=\mathbb{R}^2_+$ that is an unhappy set for the preference $\succsim_S,$ satisfies IMIA.}
\end{enumerate}{\it The statements above are related as:} (A)$\Rightarrow$(B)$\Rightarrow$(C).

\end{prop}\noi{\bf Proof} It is already shown in Remark \ref{nonc3} that (A)$\Rightarrow$(B). To prove (B)$\Rightarrow$(C), consider any $S\subseteq N$ with $|S|=2$ and suppose the induced preference $\succsim_S$ is noncompensatory. Without loss of generality (w.l.o.g.), let $S=\{1,2\}.$ Let $A$ be a closed subset of $X_S=\mathbb{R}^2_+$ that is an unhappy set for $\succsim_S$. To show $A$ satisfies IMIA (see Definition \ref{defn0}), let $x,y\in A$ with $x \td y.$ If $y>x,$ then by the monotoncity of $\succsim_S$ (which follows from the monotonicity of $\succsim$), $\wt{x}\in A$ for all $x<\wt{x}<y,$ so (i) of Definition \ref{defn0} holds. 

Next we consider $x,y\in A$ such that $y_2>x_2,$ $x_1>y_1$ (see Figure 1(a)) and show that (ii) of Definition \ref{defn0} holds. Let $x_2<\wt{x}_2<y_2$ and $\wt{x}=(x_1,\wt{x}_2).$ Using (\ref{nonc1}), for any such $\wt{x},$ we have $M(x,y)=M(\wt{x},y)=\{1\}$ and $M(y,x)=M(y,\wt{x})=\{2\}.$ Since $\succsim_S$ is noncompensatory, if $y\succ_S x$ we must have $y\succ_S \wt{x}$ and so $\wt{x}\in A$ (since $A$ is an unhappy set and $y\in A$). If $y\sim_S x$ (so that neither $y\succ_S x$ nor $x\succ_S y$), then again by the noncompensatory property of $\succsim_S,$  we must have $y\sim_S \wt{x},$ so that $\wt{x}\in A.$ 

Finally suppose $x\succ_S y.$ Suppose in contrary to (ii) of Definition \ref{defn0}, {\it for every} $\wt{x}_2$ with $x_2<\wt{x}_2<y_2$ and $\wt{x}=(x_1,\wt{x}_2),$ we have $\wt{x}\notin A.$ In that case, clearly any neighborhood $B_\ve(\wt{x})$ contains a point that is not in $A.$ Also note that $B_\ve(\wt{x})$ contains a point $z$ such that $z_1<x_1$ and $z_2>x_2,$ so that $M(x,y)=M(x,z)=\{1\}$ and $M(y,x)=M(z,x)=\{2\}.$ Since $\succsim_S$ in noncompensatory and $x\succ_S y,$ we must have $x\succ_S z,$ so $z\in A$ (as $A$ is an unhappy set and $x\in A$). Thus $B_\ve(\wt{x})$ contains at least one point in $A$ and at least one point not in $A.$ This shows $\wt{x}\in {\partial A}.$ Since $A$ is a closed set, we must have $\wt{x}\in A,$ a contradiction to the initial assertion that $\wt{x}\notin A.$ This shows that (ii) of Definition \ref{defn0} must hold.   

For the case $y_1>x_1,$ $x_2>y_2$ (Figure 1(b)), re-labeling $1,2,$ we can apply the same reasoning to show that (iii) of Definition \ref{defn0} holds. This shows (B)$\Rightarrow$(C). \qed

\vspace{0.1cm} For the preference of Example \ref{ex2}, (B) holds but (A) does not. By Proposition \ref{nonc4}, (C) also holds for that preference, so it satisfies Axiom \ref{imia}. For the preference of the following example, (C) holds, but (B) does not hold (and therefore (A) also does not hold).

\begin{ex}\label{ex50} Consider the lexi-max preference $\succsim$ (see, e.g., Bossert et al. 1994, Chistyakov and Chumakova 2018). For $x=(x_1,x_2,x_3)\in \mathbb{R}^3_+$, denote by $x^*_i$ the $i$-th highest order statistics of $x$ so that $x^*_1\geq x^*_2\geq x^*_3.$ For the lexi-max preference, for any $x_i,y_i\in \mathbb{R}_+$ we have:
(a) $x\succ y$ if either [$x^*_1>y^*_1$], or [$x^*_1=y^*_1,$ $x^*_2>y^*_2$], or [$x^*_1=y^*_1,$ $x^*_2=y^*_2,$ $x^*_3>y^*_3$] and (b) $x\sim y$ if [$x^*_1=y^*_1,$ $x^*_2=y^*_2,$ $x^*_3=y^*_3$].

\begin{figure}[!h]\centering\includegraphics[width=5.2in]{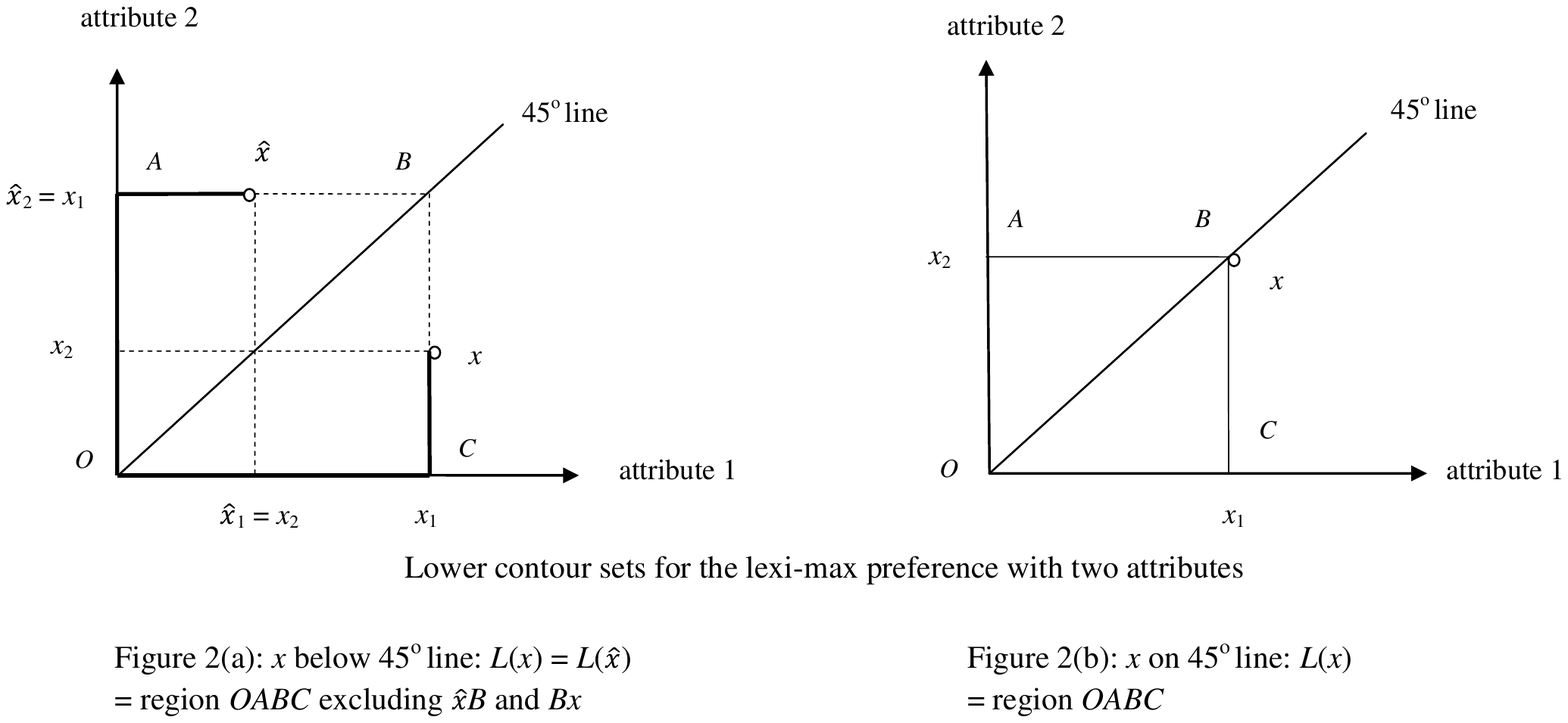}\end{figure}

Note that $\succsim$ is complete, transitive, strong monotone and for any $S\subseteq N$ with $|S|=2,$ the induced preference $\succsim_S$ is a lexi-max preference on $X_S=\mathbb{R}^2_+.$ To see $\succsim_S$ is not noncompensatory, w.l.o.g., let $S=\{1,2\}.$ The lower contour sets of the lexi-max preference on  $\mathbb{R}^2_+$ are presented in Figures 2(a)-2(b). Consider $x\in \mathbb{R}^2_+$ as in Figure 2(a). For $y$ on the line $A\hat{x}$ and $z$ on the line $\hat{x}B,$ we have $x\succ_S y$ and $z\succ_S x$ (for instance, take $x=(4,2),$ $\hat{x}=(2,4),$ $y=(1,4),$ $z=(3,4)$).  Using (\ref{nonc1}), we have $M(x,y)=M(x,z)=\{1\}$ and $M(y,x)=M(z,x)=\{2\},$ so the lexi-max preference on $\mathbb{R}^2_+$ does not satisfy noncompensation.

However, this preference satisfies Axiom \ref{imia}. To see this, w.l.o.g., let $S=\{1,2\}$ and let $A$ be any closed unhappy set for $\succsim_S$. To show $A$ satisfies IMIA (Definition \ref{defn0}), let $x,y\in \mathbb{R}^2_+$ with $x\td y.$ The conclusion is immediate for $y>x,$ so let $y_2>x_2$ and $x_1>y_1$ (by re-labeling $1,2,$ same reasoning applies when $y_1>x_1$ and $x_2>y_2$). There are two possibilities. 

(a) If $x$ is on or above the 45$^{\circ}$ line (i.e., $x_2\geq x_1$), then for any $x_2<\wt{x}_2<y_2,$ we have $\max\{x_1,\wt{x}_2\}=\wt{x}_2<y_2,$ so $y\succ_S (x_1,\wt{x}_2).$ Since $A$ is an unhappy set and $y\in A,$ we must have $(x_1,\wt{x}_2)\in A,$ so (ii) of Definition \ref{defn0} holds. 

(b) If $x$ is below the 45$^{\circ}$ line (i.e., $x_2<x_1$) as in Figure 2(a), then any $\wt{x}=(x_1,\wt{x}_2)$ with $x_2<\wt{x}_2<\min\{x_1,y_2\}$ is on the line $Bx$ and so it is a limit point of $L(x).$ Since $A$ is an unhappy set and $x\in A,$ we have $L(x)\subseteq A,$ so $\wt{x}$ is also a limit point of $A.$ As $A$ is closed, we have $\wt{x}\in A,$ which shows (ii) of Definition \ref{defn0} holds. 

\end{ex}

Proposition \ref{nonc4}, together with Examples \ref{ex2}-\ref{ex50}, clearly establishes that Axiom \ref{imia} (the IMIA requirement) is a weaker condition than Fishburn's noncompensation axiom. We can also characterize pairwise lexicographic and lexicographic preferences by replacing Axiom \ref{imia} by condition (B) of Proposition \ref{nonc3}, which is a stronger requirement than IMIA, but is still weaker than Fishburn's noncompensation axiom. 

\begin{newax}{3A}\label{conb}
For any $S\subseteq N$ with $|S|=2,$ the induced preference $\succsim_S$ is noncompensatory.
\end{newax}

\begin{cor}\label{talt}
{\it Consider any complete and transitive preference relation $\succsim$ on $X=\mathbb{R}^n_+$ where $n\geq 3.$}
\begin{enumerate}[(i)]
\item {\it The preference relation $\succsim$ is pairwise lexicographic if and only if it satisfies Axiom $\ref{mon},$ Axiom $\ref{mild}$ and Axiom $\ref{conb}.$}
\item {\it The preference relation $\succsim$ is lexicographic if and only if it satisfies Axiom $\ref{mon},$ Axiom $\ref{mild},$  Axiom $\ref{conb}$ and Axiom $\ref{irev}.$}
\end{enumerate}
\end{cor}\noi{\bf Proof} We know a pairwise lexicographic preference satisfies Axioms $\ref{mon}$-$\ref{mild}$ and a lexicographic preference satisfies Axioms $\ref{mon}$-$\ref{mild}$ and Axiom $\ref{irev}.$ The `only if' part of each of (i),(ii) follows by noting that both pairwise lexicographic and lexicographic preferences satisfy Axiom \ref{conb}. 

Noting that if Axiom \ref{conb} holds, then Axiom \ref{imia} also holds (Proposition \ref{nonc4}), the `if' part of (i) follows by Theorem \ref{tpair} and the `if' part of (ii) follows by Theorem \ref{tone}. \qed

\subsection{Fishburn's independence axiom and NRAA}

\label{nrai}

\hyperref[sec:nrai]{}

Axiom \ref{irev} (NRAA) has some resemblence with Fishburn's independence axiom (see Axiom 2 of Fishburn 1975), but they are different. The NRAA compares alternatives that are same in regard to all but two attributes (see Definition \ref{defn3}), whereas the independence axiom fixes all but one attribute: it requires $(x_i,z_{-i})\succ (y_i,z_{-i})$ if and only if $(x_i,\wt{z}_{-i})\succ (y_i,\wt{z}_{-i})$ for all $i,z_{-i},\wt{z}_{-i}.$ The following examples with $N=\{1,2,3\}$ show the independence axiom is neither necessary nor sufficient for NRAA.

\begin{ex}\label{ex1} {\bf (NRAA holds, independence axiom does not)} Consider a preference relation $\succsim$ on $X=\mathbb{R}^3_+$ that has utility function $u(x_1,x_2,x_3)=\min\{x_1,x_2\}+\min\{x_1,x_2\}x_3.$ Observe that if $S=\{1,3\}$ or $\{2,3\},$ then $u(x^S,0^{N/S})=u(y^S,0^{N/S})=0,$ so we cannot have $(x^S,0^{N/S})\succ (y^S,0^{N/S}).$ For $S=\{1,2\},$ if $(x^S,0^{N/S})\succ (y^S,0^{N/S}),$ then $\min\{x_1,x_2\}>\min\{y_1,y_2\},$ which implies that $u(x_1,x_2,z_3)>u(y_1,y_2,z_3)$ for all $z_3\geq 0.$ This shows that NRAA holds for this preference.

To see Fishburn's independence axiom does not hold, take $x_1=4,$ $y_1=1,$ $(z_2,z_3)=(2,2)$ and $(\wt{z}_2,\wt{z}_3)=(1/2,3).$ Then
$u(x_1,z_2,z_3)=6>u(y_1,z_2,z_3)=3$ so $(x_1,z_2,z_3)\succ (y_1,z_2,z_3).$ But $u(x_1,\wt{z}_2,\wt{z}_3)=u(y_1,\wt{z}_2,\wt{z}_3)=2$ so $(x_1,\wt{z}_2,\wt{z}_3)\sim (y_1,\wt{z}_2,\wt{z}_3).$
\end{ex}

\noi {\bf Independence axiom holds, NRAA does not} Consider the preference relation of Example \ref{ex2}. This preference is pairwise lexicographic, but not lexicographic. Note that $(0,6,4)\succ (0,3,8)$ (by (b) of Example \ref{ex2}), but $(1,3,8)\succ (1,6,4)$ (by (d)), so it violates Axiom \ref{irev} (NRAA).

To see whether Fishburn's independence holds, first take $i=1.$ Note from (a)-(d) of Example \ref{ex2} that $(x_1,z_2,z_3)\succ (y_1,z_2,z_3)$ if and only if $x_1>y_1.$ Then by (a), $(x_1,\wt{z}_2,\wt{z}_3)\succ (y_1,\wt{z}_2,\wt{z}_3).$ Next take $i=2.$ Note that $(z_1,x_2,z_3)\succ (z_1,y_2,z_3)$ if and only if $x_2>y_2$ (by (b),(d) of Example \ref{ex2}) which implies $(\wt{z}_1,x_2,\wt{z}_3)\succ (\wt{z}_1,y_2,\wt{z}_3).$ Finally for $i=3,$ $(z_1,z_2,x_3)\succ (z_1,z_2,y_3)$ if and only if $x_3>y_3$ (by (c),(d)), which implies $(x_1,\wt{z}_2,\wt{z}_3)\succ (y_1,\wt{z}_2,\wt{z}_3).$ This shows Fishburn's independence axiom holds for this example.

\subsection{Mild continuity of induced preferences}

\label{mildc}

\hyperref[sec:mildc]{}

Note that Axiom \ref{mild} requires that the induced preference $\succsim_S$ is mildly continuous for any $S$ that contains only two attributes. It does not require the preference $\succsim$ to be mildly continuous. For a lexicographic preference, any such induced preference, as well as the preference itself are mildly continuous. But in general, mild continuity of an induced preference is neither necessary nor sufficient for a preference to be mildly continuous. This is demonstrated in the following examples with $N=\{1,2,3\}.$

\begin{ex}\label{ex0} {\bf (induced preference mildly continous, but not preference)} Consider a preference relation $\succsim$ on $X=\mathbb{R}^3_+.$ First note that for $x,y\in X,$ either (A) $x_i=y_i=0$ for one or more $i=1,2,3$ or (B) at least one of $x_i,y_i$ is positive for all $i=1,2,3.$ Note that (A),(B) are mutually exclusive and exhaustive. 

(A) If $x,y\in X$ are such that $x_i=y_i=0$ for one or more $i,$ the following hold: 

\noi (i) $(0,x_2,x_3)\succ (0,y_2,y_3)$ if $x_2+x_3>y_2+y_3,$ $(0,x_2,x_3)\sim (0,y_2,y_3)$ if $x_2+x_3=y_2+y_3;$ (ii) $(x_1,0,x_3)\succ (y_1,0,y_3)$ if $x_1+x_3>y_1+y_3,$ $(x_1,0,x_3)\sim (y_1,0,y_3)$ if $x_1+x_3=y_1+y_3;$ (iii) $(x_1,x_2,0)\succ (y_1,y_2,0)$ if $x_1+x_2>y_1+y_2,$ $(x_1,x_2,0)\sim (y_1,y_2,0)$ if $x_1+x_2=y_1+y_2.$ By (i)-(iii), for any $S\subseteq N$ with $|S|=2,$ the induced preference $\succsim_S$ is continuous, so $\succsim_S$ is mildly continuous.

(B) If $x,y\in X$ are such that for all $i,$ at least one of $x_i,y_i$ is positive, the following hold where  $\ul{x}=\min\{x_1,x_2,x_3\}$: 

(i) if $\ul{x}>\ul{y}=0,$ then $x\succ y$ and if $\ul{x}=\ul{y}=0,$ then $x\sim y;$ (ii) if $\ul{x},\ul{y}$ are both positive, then  $x\succ y$ if $x_1+x_2+x_3>y_1+y_2+y_3$ and $x\sim y$ if $x_1+x_2+x_3=y_1+y_2+y_3.$

Let $x=(2,1,1),$ $y=(0,2,2).$ Note that $x\td y,$ $\ul{x}=1>\ul{y}=0.$ By (B)(i), $x\succ y.$ Consider any $0<\ve<4$ and let $z=(\ve/2,2,2).$ Then $z\in B_\ve(y).$ Since $\ul{z}=\ve/2>0$ and $z_1+z_2+z_3=4+\ve/2>x_1+x_2+x_3=4,$ by (B)(ii), $z\succ x.$ This shows $\succsim$ is not mildly continuous (see Definition \ref{mcont}), although for every $S\subseteq N$ with $|S|=2,$ the induced preference $\succsim_S$ is mildly continuous.

\end{ex}

\begin{ex}\label{ex01} {\bf (preference mildly continous, but not induced preference)}  Consider a preference relation $\succsim$ on $X=\mathbb{R}^3_+$ that has the following properties, where $\ul{x}^S=\min_{i\in S}x_i$ for any non empty $S\subseteq N.$ 

(A) If $x,y\in X$ are such that $x_i=y_i=0$ for one or more $i,$ the following hold:

for any $S\subseteq N$ with $|S|=2$: (i) if $\ul{x}^S>\ul{y}^S=0,$ then $(x^S,0^{N/S})\succ (y^S,0^{N/S})$ and if $\ul{x}^S=\ul{y}^S=0,$ then $(x^S,0^{N/S})\sim (y^S,0^{N/S});$ (ii) if $\ul{x}^S,$ $\ul{y}^S$ are both positive, then  $(x^S,0^{N/S})\succ (y^S,0^{N/S})$ if $\sum_{i\in S}x_i>\sum_{i\in S}y_i$ and $(x^S,0^{N/S})\sim (y^S,0^{N/S})$ if $\sum_{i\in S}x_i=\sum_{i\in S}y_i.$

Note that the induced preference $\succsim_S$ is not mildly continuous for any $S\subseteq N$ with $|S|=2.$ We show this for $S=\{1,2\}$ (similar reasoning applies for other subsets). Let $x^S=(1,1),$ $y^S=(0,2).$ Note that $x^S\td y^S,$ $\ul{x}^S=1>\ul{y}^S=0.$ By (A)(i), $x^S\succ_S y^S.$ Consider any $0<\ve<4$ and let $z^S=(\ve/2,2).$ Note that $z^S\in B_\ve(y^S).$ Since $\ul{z}^S=\ve/2>0$ and $z_1+z_2=2+\ve/2>x_1+x_2=2,$ by (A)(ii), $z^S\succ x^S.$ This shows the induced preference $\succsim_S$ is not mildly continuous. 

(B) If $x,y\in X$ are such that at least one of $x_i,y_i$ is positive for all $i,$ the following hold:

$x\succ y$ if $x_1+x_2+x_3>y_1+y_2+y_3$ and $x\sim y$ if $x_1+x_2+x_3=y_1+y_2+y_3.$

Note that if $x\td y,$ at least one of $x_i,y_i$ must be positive for all $i.$ Consider any $x\td y$ such that $x\succ y.$ Then we must have $x_1+x_2+x_3>y_1+y_2+y_3.$ We can find sufficiently small $\ve>0$ such that for all $\wt{x}\in B_\ve(x),$ $\wt{y}\in B_\ve(y)$: $\wt{x}\td \wt{y}$ (implying at least one of $\wt{x}_i,\wt{y}_i$ is positive for all $i$)  and $\wt{x}_1+\wt{x}_2+\wt{x}_3>\wt{y}_1+\wt{y}_2+\wt{y}_3,$ so by (B), $\wt{x}\succ \wt{y}.$ This shows $\succsim$ is mildly continuous (see Definition \ref{mcont}), although the induced preference $\succsim_S$ is not mildly continuous for any $S\subseteq N$ with $|S|=2.$   

\end{ex}

\subsection{Robustness of the axioms}

Finally we give examples of complete and transitive preferences to show that if one of the four Axioms \ref{mon}-\ref{irev} does not hold, we do not get a lexicographic preference. 

\vspace{0.1cm} \noi {\bf All axioms except Axiom \ref{irev} hold} The preference of Example \ref{ex2} is pairwise lexicographic but not lexicographic. By Theorem \ref{tpair}, Axioms \ref{mon}-\ref{imia} hold for this preference, but we have already seen Axiom \ref{irev} does not hold for this preference. 

\begin{ex}\label{ex4} \rm {\bf (All axioms except Axiom \ref{mon} hold)} Consider the preference relation $\succsim$ on $X=\mathbb{R}^3_+$ that has utility function $u(x_1,x_2,x_3)=x_1$ (this is a dominant preference where attribute $1$ is the dominant attribute). Note that the induced preference $\succsim_S$ is not strong monotone for $S=\{1,2\},$ so Axiom \ref{mon} is violated. We show the remaining three axioms hold.

Note that for any $S\subseteq N$ with $|S|=2,$ $\succsim_S$ is continuous, so Axiom \ref{mild} holds.  As $\succsim_S$ is continuous, any closed unhappy set for  $\succsim_S$ is a lower contour set (Corollary \ref{cor1}(iii)). To see Axiom \ref{imia} holds, first let $S=\{1,2\}.$ Since $x^S=(x_1,x_2)\succ_S y^S=(y_1,y_2)$ if and only if $x_1>y_1,$ the lower contour set of $x^S$ under $\succsim_S$ is $L^*(x^S)=\{y^S\in \mathbb{R}^2_+|0\leq y_1\leq x_1\}.$ To see IMIA (see Definition \ref{defn0}) holds for $L^*(x^S),$ consider any $y,z\in L^*(x^S)$ such that $y\td z.$ If $y>z,$ then $z_1<y_1\leq x_1,$ so any $\wt{z}$ with $y>\wt{z}>z$ has $\wt{z}_1<x_1,$ which shows $\wt{z}\in L^*(x^S).$ If $y_2>z_2$ and $z_1>y_1,$ then $y_1<z_1\leq x_1,$ so $(z_1,\wt{z}_2)\in L^*(x^S)$ for any $z_2<\wt{z}_2<y_2.$ If $y_1>z_1$ and $z_2>y_2,$ then $z_1<y_1\leq x_1.$ So $\wt{z}_1< x_1$ for any $z_1<\wt{z}_1<y_1,$ which shows $(\wt{z}_1,z_2)\in L^*(x^S)$ for any such $\wt{z}_1.$ This shows IMIA holds for any closed unhappy set for $\succsim_S$ when $S=\{1,2\}.$  Similar reasoning applies when  $S=\{1,3\}.$ Finally for $S=\{2,3\},$ for any $x^S,y^S\in X_S$ we have $x^S\sim_S y^S$ so the only non empty unhappy set for $\succsim_S$ if $X_S=\mathbb{R}^2_+$ and for it, IMIA trivially holds.

To see Axiom \ref{irev} holds, note that if $(x_1,x_2,0)\succ (y_1,y_2,0),$ then $x_1>y_1$ and $(x_1,x_2,z_3)\succ (y_1,y_2,z_3)$ for any $z_3.$ This shows NRAA (see Definition \ref{defn3}) holds when $S=\{1,2\}.$ Similar reasoning applies when  $S=\{1,3\}.$ Finally when $S=\{2,3\},$ for any $x^S=(x_2,x_3),y^S=(y_2,y_3)\in X_S,$ we have $(0,x_2,x_3)\sim_S (0,y_2,y_3),$ so NRAA is satisfied vacuously.  This shows both Axiom \ref{imia} and Axiom \ref{irev} hold. \end{ex}

\begin{ex}\label{ex3} \rm {\bf (All axioms except Axiom \ref{imia} hold)} Consider the preference relation $\succsim$ on $X=\mathbb{R}^3_+$ with utility function $u(x)=x_1+x_2+x_3$ (perfect substitutes). Since $\succsim$ is strong monotone, so is $\succsim_S$ for any non empty $S\subseteq N,$ so Axiom \ref{mon} holds. Since it is continuous, Axiom \ref{mild} also holds. Since the utility function is additively separable, Axiom \ref{irev} (NRAA) also holds.

However, Axiom \ref{imia} does not hold. For instance, let $S=\{1,2\}.$ Consider the induced preference $\succsim_S,$ which has utility function $u_S(x^S)=x_1+x_2.$ It is continuous, so by Corollary \ref{cor1}(iii), any closed unhappy set of $\succsim_S$ is a lower contour set. Consider the lower contour set $L^*(x^S)$ for $x^S=(1,3).$ Let $y^S=(3,1).$ Then $x^S,y^S\in L^*(x^S).$ Also note that $x^S\td y^S$ with $y_1>x_1$ and $x_2>y_2$ (Figure 1(b)). But there is no $\wt{x}_1$ with $x_1<\wt{x}_1<y_1$ such that $(\wt{x}_1,x_2)\in L^*(x^S).$ This is because for any $\wt{x}_1>x_1,$ we have $\wt{x}_1+x_2>x_1+x_2=u_S(x^S),$ so such a point will be outside $L^*(x^S).$
\end{ex}

\noi {\bf All axioms except Axiom \ref{mild} hold} Consider the lexi-max preference $\succsim$ given in Example \ref{ex50}. Since $\succsim$ is strong monotone, so is $\succsim_S$ is strong monotone for any non empty $S\subseteq N,$ so Axiom \ref{mon} holds. We have already shown in Example \ref{ex50} that Axiom \ref{imia} holds. 

To verify Axiom \ref{irev} (NRAA) holds, consider any $S\subseteq N$ with $|S|=2.$ Let $S=\{1,2\}$ (same reasoning applies for other subsets). Suppose $(x_1,x_2,0)\succ (y_1,y_2,0).$ Then either [$x^*_1>y^*_1$] or [$x^*_1=y^*_1,$ $x^*_2>y^*_2$]. Take any $x_3\in \mathbb{R}_+.$ Let $a,b\in \mathbb{R}^3_+$ be as follows:
$$a_1=x_1,a_2=x_2,a_3=x_3\mbox{ and }b_1=y_1,b_2=y_2,b_3=x_3$$To prove NRAA holds, we have to show $a\succ b.$ We have the following possibilities. 

(i) $x_3\geq x^*_1$: Then
$a^*_1=b^*_1=x_3,$ $a^*_2=x^*_1,$ $b^*_2=y^*_1,$ $a^*_3=x^*_2$ and $b^*_3=y^*_2.$ Since either [$a^*_2>b^*_2$] or [$a^*_2=b^*_2,$ $a^*_3>b^*_3$], we conclude that $a\succ b.$

(ii) $x_3<x^*_1$ and $x^*_1>y^*_1$: Then $a^*_1=x^*_1>b^*_1=\max\{y^*_1,x_3\},$ so we have $a\succ b.$

(iii) $x_3<x^*_1,$ $x^*_1=y^*_1$ and $x^*_2>y^*_2$: Then $a^*_1=b^*_1=x^*_1.$ If $x^*_2\leq x_3,$ then $a^*_2=b^*_2=x_3$ and $a^*_3=x^*_2>b^*_3=y^*_2,$ so we have $a\succ b.$ If $x^*_2>x_3,$ then $a^*_2=x^*_2$ and $b^*_2=\max\{y^*_2,x_3\}<a^*_2,$ so we have $a\succ b.$ This shows that  Axiom \ref{irev} also holds.

To see Axiom \ref{mild} does not hold for this preference, let $S=\{1,2\}$ and consider the induced preference $\succsim_S.$ Take $x\in \mathbb{R}^2_+$ such that $x$ is below the 45$^\circ$ line as in Figure 2(a). Take any $y$ on the line $A\hat{x}$ (for instance, take $x=(4,2),$ $\hat{x}=(2,4),$ $y=(1,4)$). Note that $x\td y$ and $y\in L(x),$ so $x\succ_S y.$ However, any neighborhood of $y$ contains points outside $L(x),$ which shows $\succsim_S$ is not mildly continuous.

\begin{rem} Finally we note that without both completeness and transitivity, we do not get lexicographic preference even with all Axioms \ref{mon}-\ref{irev}. For $x,y\in X,$ we say $x,y$ are {\it non-comparable}, denoted by $x\bowtie y,$ if $\exists$ $i,j$ such that $y_i>x_i$ and $x_j>y_j.$ Consider a preference relation $\succsim$ on $X=\mathbb{R}^3_+$ that is reflexive and strong monotone. If $x\sim y$ whenever $x\bowtie y,$ the preference relation is complete but not transitive and satisfies Axioms \ref{mon}-\ref{irev}. If whenever $x\bowtie y,$ we have neither $x\succsim y$ nor $y\succsim x,$ the preference relation is transitive but not complete and satisfies Axioms \ref{mon}-\ref{irev}. 
\end{rem}

\section*{Acknowledgements}

We express our sincere gratitude to the editor, an anonymous associate editor and two anonymous referees for their helpful comments and suggestions. For their comments, we also thank the seminar participants at 13th Annual Conference on Economic Growth and Development at Indian Statistical Institute, New Delhi and XXVIIth Annual General Conference on Contemporary Issues in Development Economics at Jadavpur University. Sen gratefully acknowledges research support by the Department of Economics, Ryerson University.

\section*{Appendix}
\label{app}

\hyperref[sec:app]{}

\noi {\bf Proof of Proposition \ref{propone}} (i) Suppose there are two unhappy sets $A,B$ and $x\in A,y\in B$ such that $x\notin B,y\notin A.$ By definition of unhappy sets we must have $x\succ y$ and $y\succ x,$ a contradiction.

\vspace{0.1cm} For (ii), by Result \ref{res1} we know there exists $x\in {\partial A}.$

\vspace{0.1cm} (ii)(a) If the assertion is not true, then by completeness one of $x,y$ is strictly preferred to the other. Without loss of generality, let $x\succ y.$ Then mild continuity implies $\exists$ $\ve>0$ such that all points in $B_{\ve}(x)$ is strictly preferred to all points in $B_{\ve}(y).$ Since $x,y\in {\partial A},$ $\exists$ $\wt{x}\in B_{\ve}(x),$ $\wt{y}\in B_{\ve}(y)$ such that $\wt{x}\in A,$ $\wt{y}\notin A$ and we have $\wt{x}\succ \wt{y}.$ A contradiction since $A$ is an unhappy set.

\vspace{0.1cm} (ii)(b) If the assertion is not true, then $y\succ x$ and mild continuity implies $\exists$ $\ve>0$ such that $y$ is strictly preferred to all points in $B_{\ve}(x).$ Since $x\in {\partial A},$ $\exists$ $\wt{x}\in B_{\ve}(x)$ such that $\wt{x}\notin A$ and we have $y\succ \wt{x},$ a contradiction since $y\in A$ and $A$ is an unhappy set.

\vspace{0.1cm} (ii)(c) As $x\succ y,$ mild continuity implies $\exists$ $\ve>0$ such that all points in $B_{\ve}(x)$ is strictly preferred to $y.$ Since $x\in {\partial A},$ $\exists$ $\wt{x}\in B_{\ve}(x)$ such that $\wt{x}\in A$ and we have $\wt{x}\succ y.$ Since $A$ is an unhappy set, we must have $y\in A.$ \qed

\vspace{0.2cm} \noi {\bf Proof of Corollary \ref{cor1}} (i)-(ii) As the preference relation is continuous, strict preference orders are preserved around small neighborhoods of {\it any two} points rather than only totally different points. Thus for any $x\in \partial{A},$ conclusions of Proposition \ref{propone}(ii) hold for {\it any} $y\in X.$ Part (i) of the corollary follows by applying Proposition \ref{propone}(ii)(a) for any $y\in X.$ Part (ii) follows by applying Proposition \ref{propone}(ii)(b)-(c) for any $y\in X.$

(iii) By Result \ref{res1}, the set $A$ cannot be both open and closed, so we can have either (a) $A$ is neither open nor closed, or (b) $A$ is either open or closed, but not both. First we rule out (a). To see this, suppose there is an unhappy set $A$ that is neither open nor closed. Since $A$ is not open, $\exists$ $x\in A$ such that every neighborhood of $x$ contains a point outside $A,$ so we have $x\in \partial{A}.$
Since $A$ is not closed, $\exists$ $y\notin A$ which is a limit point of $A,$ that is, every neighborhood of $y$ contains a point in $A,$ so we have $y\in \partial{A}.$ Since $x,y\in \partial{A},$ by (i) we have $x\sim y.$ However, since $x\in A,$ $y\notin A$ and $A$ is an unhappy set, we must have $y\succ x,$ a contradiction. This rules out (a). So the set $A$ is either open or closed, but not both.

Suppose $A$ is open. Let $x\in \partial{A}.$ We must have $x\notin A$ since every neighborhood of $x$ contains a point outside $A.$ As $A$ is an unhappy set, if $y\sim x,$ we must have $y\notin A.$ This shows $A\cap I(x)=\emp.$ Since $A\subseteq L(x)=\ul{L}(x)\cup I(x)$ (by (ii)), we must have $A\subseteq \ul{L}(x)$ and again by (ii) we have $A=\ul{L}(x).$

Suppose $A$ is closed. Let $x\in \partial{A}.$ As $A$ is closed, we have $x\in A.$ Since $A$ is an unhappy set, if $y\sim x,$ we must have $y\in A.$ This shows $I(x)\subseteq A.$ Since $\ul{L}(x)\subseteq A,$ (by (ii)), we conclude $L(x)=\ul{L}(x)\cup I(x)\subseteq A.$ Again by (ii) it follows that $A={L}(x).$ \qed

\vspace{0.2cm} \noi {\bf Proof of Proposition \ref{proptwo}} Note by the definition of a lower contour set that any point in $X\sm \ol{L}(x)$ is strictly preferred to any point in $L(x).$ To prove $\ol{L}(x)$ is an unhappy set, it remains to show that if $a\in X\sm \ol{L}(x)$ and $b\in \partial{L(x)},$ then $a\succ b.$

Since $\succsim$ is complete, if the result does not hold, there are $a\in X\sm \ol{L}(x),$ $b\in \partial{L(x)}$ such that $b\succsim a.$ Denote
$$E=\{i\in N|a_i=b_i=0\}\mbox{ and }S=N\sm E$$so that $S\cup E=N.$ Since $\succsim$ is strong monotone, $0^N\in L(x).$ So $a\neq 0^N.$ Thus $S\neq\emp $ and there is at least one $i\in S$ with $a_i>0.$

Since $a\in X\sm L(x)$ and $a\notin \partial L(x),$ there is a neighborhood $B_\ve(a)\subseteq X\sm L(x).$ We can construct $\wt{a}\in B_\ve(a)$ (so that $\wt{a}\in X\sm L(x)$) such that (i) if $i\in E,$ then $\wt{a}_i=a_i=0;$ (ii) if $i\in S$ and $a_i=0,$ then $\wt{a}_i=0;$ and (iii) if $i\in S$ and $a_i>0,$ then $0<\wt{a}_i<a_i$ and $\wt{a}_i\neq b_i.$ Note that $\wt{a}^{S}\td b^{S}.$ By strong monotonicity of $\succsim,$ $a\succ \wt{a}$ and transitivity implies $b\succ \wt{a}.$

If $E=\emp,$ then $N=S.$ So $\wt{a}=\wt{a}^{S},$ $b=b^{S}$ and we have $\wt{a}\td b.$ Since $L(x)$ is an unhappy set, $b\in \partial{L(x)}$ and  $b\succ \wt{a},$ by Proposition \ref{propone}(ii)(c) we must have $\wt{a}\in L(x)$ which is a contradiction since $\wt{a}\in X\sm L(x).$

If $E\neq \emp,$ then $\wt{a}=(\wt{a}^{S},0^E)$ and $b=(b^{S},0^E).$ Since $b\in {\partial L(x)},$ every neighborhood $B_\ve(b)$ contains a point $\bar{b}\in L(x).$ From such $\bar{b},$ construct  $c$ as $c=(\bar{b}^{S},0^E).$ By monotonicity of $\succsim,$ we have $\bar{b}\succsim c.$ Hence $c\in L(x).$ Moreover $d(b^{S},\bar{b}^{S})=d(b,c)\leq d(b,\bar{b})<\ve,$ so $\bar{b}^{S}\in B_\ve(b^{S}).$  This shows for every neighborhood $B_\ve(b^S)$ of $b^S$:
\be\label{0}\exists\mbox{ }\bar{b}^S\in B_\ve(b^S)\mbox{ such that }(\bar{b}^{S},0^E)\in L(x)\ee Consider the induced preference $\succsim_S$ on $X_S.$ Note that for $y^S,z^S\in X_S:$
\be\label{eq}y^S\succsim_S z^S\Leftrightarrow (y^S,0^{E})\succsim (z^S,0^{E})\ee It is given $\succsim_S$ is mildly continuous on $X_S.$
Since $b=(b^S,0^E)\succ \wt{a}=(\wt{a}^S,0^E),$ by (\ref{eq}) we have $b^S\succ_S \wt{a}^{S}.$ Since $b^S\td\wt{a}^S,$ by mild continuity of $\succsim_S,$ there is a neighborhood $B_\ve(b^{S})$ such that for every $\wt{b}^S\in B_\ve(b^{S})$ we have $\wt{b}^{S}\succ_S \wt{a}^{S},$ so by (\ref{eq})
\be\label{1}(\wt{b}^{S},0^E)\succ (\wt{a}^{S},0^E)=\wt{a}\mbox{ for all }\wt{b}^S\in B_\ve(b^{S})\ee By (\ref{0}) and (\ref{1}), it follows that $\wt{a}\in L(x),$ which is a contradiction since $\wt{a}\in X\sm L(x).$ \qed

\vspace{0.2cm} \noi {\bf Proof of Lemma \ref{leximia}} Suppose the set of attributes is $N=\{1,2\}.$ Let $\succsim$ be a lexicographic preference on $X=\mathbb{R}^2_+$ and suppose $1$ is the most important attribute of $\succsim.$

Let $A\subseteq X$ be a closed unhappy set for $\succsim$ and $x,y\in A$ such that $x\td y.$ 
If $y>x,$ then clearly (i) of Definition \ref{defn0} holds by monotonicity of $\succsim.$ If $y_1>x_1$ and $x_2>y_2$ (Figure 1(b)), then for any $x_1<\wt{x}_1<y_1,$ we have $y\succ (\wt{x}_1,x_2).$ As $A$ is an unhappy set and $y\in A,$ we have $(\wt{x}_1,x_2)\in A,$ so (iii) of Definition \ref{defn0} holds.

Finally let $y_2>x_2$ and $x_1>y_1$ (Figure 1(a)). Consider any $\wt{x}_2>x_2$ and let $\wt{x}=(x_1,\wt{x}_2).$ Any neighborhood of $\wt{x}$ contains a point $z$ such that $z_1<x_1.$ So $x\succ z.$ As $A$ is an unhappy set, we have $z\in A.$ Thus any neighborhood of $\wt{x}$ contains a point $z\neq\wt{x}$ such that $z\in A,$ so $\wt{x}$ is a limit point of $A.$ As $A$ is closed, $\wt{x}\in A.$ This shows $\wt{x}=(x_1,\wt{x}_2)\in A$ for any $\wt{x}_2>x_2,$ so (ii) of Definition \ref{defn0} holds. \qed 

\vspace{0.2cm}\noi {\bf Boundedness of unhappy sets under IMIA} For $X=\mathbb{R}^2_+,$ when an unhappy set satisfies IMIA (Definition \ref{defn0}), there are some useful implications with respect to boundedness of that set which are presented in Lemma \ref{propthree}. This result will be used to prove Lemma \ref{bound}. Let $X_1=\mathbb{R}_+,$ $X_2=\mathbb{R}_+,$ $A\subseteq X=\mathbb{R}^2_+$ and $x=(x_1,x_2)\in A.$ Denote 
$$A(x_2):=\{y_1\in X_1|(y_1,x_2)\in A\}\mbox{ and }A(x_1):=\{y_2\in X_2|(x_1,y_2)\in A\}$$For $i,j=1,2,$ $i\neq j,$  we say $A(x_j)$ is bounded above if there is a number $k$ such that $y_i\leq k$ for all $y_i\in A(x_j).$

\begin{newlem}{A1}\label{propthree}
{\it Consider a complete, transitive and strong monotone preference relation $\succsim$ on $X=\mathbb{R}^2_+$. Let $A\subseteq X$ be a closed unhappy set that satisfies IMIA and let $x\in A.$}

\begin{enumerate}[(i)]

\item {\it For $i,j=1,2$ and $i\neq j,$ if $A(x_j)$ is bounded above, then}

\begin{enumerate}[(a)]

\item {\it $\exists$ $\alpha(x_j)$ $($the least upper bound of $A(x_j))$ such that $A(x_j)=[0,\alpha(x_j)].$}

\item {\it For any $z\in A,$ $A(z_j)$ is also bounded above with the same least upper bound. That is, $\alpha(x_j)=\alpha(z_j)=\alpha_i$ and
$A(x_j)=A(z_j)=[0,\alpha_i].$ Moreover $(\alpha_i,x_j),(\alpha_i,z_j)$ are both boundary points of $A.$}

\end{enumerate}

\item {\it If $A(x_j)$ is not bounded above, then for any $z\in A,$ $A(z_j)$ is also not bounded above and $A(x_j)=A(z_j)=\mathbb{R}_+.$}

\item {\it If $\succsim$ is mildly continuous, then there can be at most one $i=1,2$ such that the following hold $($where $j=2$ if $i=1$ and $j=1$ if $i=2)$}: 
{\it \be\label{p1}\mbox{ there exists }z\in A\mbox{ with }z>(0,0)\mbox{ such that }A(z_j)\mbox{ is bounded above }\ee}
\end{enumerate}\end{newlem}\noi{\bf Proof} For (i)-(ii), without loss of generality, let $i=1$ and $j=2.$ 

(i)(a) Since $x=(x_1,x_2)\in A,$ we have $x_1\in A(x_2),$ so $A(x_2)$ is a non empty subset of $\mathbb{R}.$ As $A(x_2)$ is bounded above, by the least-upper-bound property of $\mathbb{R}$ (see Theorem 1.19, Rudin, 1976) $A(x_2)$ has a least upper bound $\alpha(x_2)\geq x_1.$ 

If $\alpha(x_2)=0,$ then it must be the only point of $A(x_2)$ and the result is immediate since $A(x_2)=\{0\}=[0,\alpha(x_2)].$ So let $\alpha(x_2)>0.$ As $A$ is an unhappy set, strong monotonicity of the preference implies if $y_1\in A(x_2),$ then $\wt{y}_1\in A(x_2)$ for any $\wt{y}_1<y_1.$ So $(y_1,x_2)\in A$ whenever $0\leq y_1<\alpha(x_2)$ and $(y_1,x_2)\notin A$ whenever  $y_1>\alpha(x_2).$ Finally note that any neighborhood of $(\alpha(x_2),x_2)$ contains a point $(y_1,x_2)$ with $y_1<\alpha(x_2).$ So $(\alpha(x_2),x_2)$ is a limit point of $A.$ Since $A$ is a closed set, $(\alpha(x_2),x_2)\in A.$ This proves $A(x_2)=[0,\alpha(x_2)].$

\vspace{0.1cm}(i)(b) Since $z\in A,$ we have $z_1\in A(z_2),$ so $A(z_2)$ is non empty. First suppose $z_2\geq x_2.$ Since $A$ is an unhappy set and $(\alpha(x_2),x_2)\in A,$ if $(y_1,z_2)\in A$ with $y_1>\alpha(x_2),$ then by strong monotonicity  $(\wt{x}_1,x_2)\in A$ for any $\alpha(x_2)<\wt{x}_1<y_1.$ This implies $\wt{x}_1\in A(x_2),$ contradicting (i)(a). Next suppose $z_2<x_2.$ Since $A$ is an unhappy set satisying IMIA and $(\alpha(x_2),x_2)\in A,$ if $(y_1,z_2)\in A$ with $y_1>\alpha(x_2),$ then by Definition \ref{defn0}(iii), $\exists$ $\alpha(x_2)<\wt{x}_1<y_1$ such that $(\wt{x}_1,x_2)\in A,$ implying $\wt{x}_1\in A(x_2),$ again contradicting (i)(a). This shows $A(z_2)$ must be bounded above by $\alpha(x_2)$ and it has a least upper bound $\alpha(z_2)\leq \alpha(x_2).$ 

Applying the result of (i)(a) for $A(z_2),$ we have $A(z_2)=[0,\alpha(z_2)]$ and $(\alpha(z_2),z_2)\in A.$ Switching the roles of $x_2$ and $z_2$ in the last paragraph, it follows that $\alpha(x_2)\leq \alpha(z_2).$ This shows $\alpha(z_2)=\alpha(x_2).$ Denoting their common value by $\alpha_1,$ we have $A(x_2)=A(z_2)=[0,\alpha_1].$ 

As $(\alpha_1,x_2)\in A$ and any neighborhood of $(\alpha_1,x_2)$ contains a point $y=(y_1,x_2)$ such that  $y_1>\alpha_1,$ we have $y\in X\sm A.$ This shows $(\alpha_1,x_2)\in \partial{A}.$ The same holds for $(\alpha_1,z_2).$

\vspace{0.1cm} (ii) As $A(x_2)$ is not bounded above, for any $k>0,$ there is $y_1>k$ such that $y_1\in A(x_2).$ As $A$ is an unhappy set, strong monotonicity of the preference implies all $\wt{y}_1<y_1$ also belongs to $A(x_2).$ This shows $A(x_2)=\mathbb{R}_+.$

\vspace{0.1cm} For $z\in A,$ if the set $A(z_2)$ is bounded above, it has a least upper bound $\alpha(z_2)$ and by (i)(a), 
$A(z_2)=[0,\alpha(z_2)].$ Then by (i)(b), the set $A(x_2)$ is also bounded above and $A(x_2)=[0,\alpha(z_2)],$ a contradiction. This shows $A(z_2)$ must be also not bounded above and $A(z_2)=\mathbb{R}_+.$

\vspace{0.1cm} (iii) We have to show that there can be at most one $i$ such that (\ref{p1}) holds (where $j=2$ if $i=1$ and $j=1$ if $i=2$). Suppose on the contrary, (\ref{p1}) holds for both $i=1,2.$  Then there are $y,z\in A$ with $y>(0,0),$ $z>(0,0)$ such that $A(y_2),$ $A(z_1)$ are both bounded above. By part (i), there are numbers $\alpha_1\geq y_1,$ $\alpha_2\geq z_2$ such that $A(y_2)=[0,\alpha_1]$ and $A(z_1)=[0,\alpha_2].$

Since $y>(0,0)$ and $z>(0,0),$ we can construct $\wt{x},\wt{y}$ such that $0<\wt{y}_i<\wt{x}_i<\min\{y_i,z_i\}$ for $i=1,2.$ Then $y>\wt{x}>\wt{y}.$ As $A$ is an unhappy set and $y\in A,$ by strong monotonicity, $\wt{x},\wt{y}\in A.$ Then by part (i)(b), $A(\wt{x}_2)=A(\wt{y}_2)=[0,\alpha_1].$ Let $\ol{x}=(\alpha_1,\wt{x}_2),\ol{y}=(\alpha_1,\wt{y}_2).$ By (i)(b), $\ol{x},\ol{y}\in \partial{A}.$

Since $z>(0,0),$ we can construct $\wt{z}$ such that $0<\wt{z}_1<\min\{z_1,\alpha_1\}$ and $\wt{z}_2=z_2.$ Then $z\geq \wt{z}.$ As $z\in A,$ by strong monotonicity, $\wt{z}\in A$ and by part (i)(b), $A(\wt{z}_1)=[0,\alpha_2].$ Let $\ol{z}=(\wt{z}_1,\alpha_2).$ By part (i)(b), $\ol{z}\in \partial{A}.$

Note that (a) $\ol{y}_1=\ol{x}_1=\alpha_1>\ol{z}_1$ and (b) $\ol{y}_2<\ol{x}_2<z_2\leq \alpha_2=\ol{z}_2.$ This shows $\ol{x}\td\ol{z}$ and $\ol{y}\td\ol{z}.$ As $\ol{x},\ol{y},\ol{z}$ are all boundary points of an unhappy set $A$ of a mildly continuous preference, by Proposition \ref{propone}(ii)(a): $\ol{x}\sim \ol{z}$ and $\ol{y}\sim \ol{z}.$ Then transitivity implies $\ol{x}\sim \ol{y}.$ However, since $\ol{x}_1=\ol{y}_1$ and $\ol{x}_2>\ol{y}_2,$ by strong monotonicity of the preference relation we have $\ol{x}\succ \ol{y},$ which is a contradiction. This proves that there can be at most one $i=1,2$ where (\ref{p1}) holds. \qed

\vspace{0.2cm} \noi {\bf Proof of Lemma \ref{bound}} Let $A\in \mathcal{A}_\succsim.$ By (\ref{a}), we know $\exists$ $z=(z_1,z_2)\in A$ with $z>(0,0).$ By Lemma \ref{propthree} there can be at most one $i=1,2$ where (\ref{p1}) holds. Suppose (\ref{p1}) fails to hold for both $i=1,2.$ Then $A(z_2)$ is unbounded, so $(x_1,z_2)\in A$ for any $x_1.$ Also $A(z_1)$ is unbounded, so by Lemma \ref{propthree}(ii) $A(x_1)$ is also unbounded, implying $(x_1,x_2)\in A$ for any $x_2.$ This shows any  $x=(x_1,x_2)\in A,$ so $A=X.$ This is a contradiction since by (\ref{a}), $A$ is a proper subset of $X.$  
This shows there is exactly one $i=1,2$ for which (\ref{p1}) holds. Denote this $i$ by $i^*.$ 

Without loss of generality, let $i^*=1.$ Then for any $z\in A,$ $A(z_2)$ is bounded above and $A(z_1)$ is unbounded. By Lemma \ref{propthree}(i)-(ii), $\exists$ $\alpha^A$ (which is positive since by (\ref{a}) there is a $z\in A$ with $z_1>0$) such that for any $z\in A$: (a) $A(z_2)=[0,\alpha^A]$ and (b) $A(z_1)=\mathbb{R}_+.$ 

For any $z=(z_1,z_2)\in A,$ by (a), $(\alpha^A,z_2)\in A.$ Then (b) implies $(\alpha^A,y_2)\in A$ 
for any $y_2\in \mathbb{R}_+.$ Since $A$ is an unhappy set, strong monotonicity of the preference implies that $(y_1  ,y_2)\in A$ for any $y_1\in [0,\alpha^A]$ and $y_2\in \mathbb{R}_+.$ This shows $\{y\in X|0\leq y_1 \leq \alpha^A\}\subseteq A.$
Note from (a) that if $y\in X$ has $y_1 >\alpha^A,$ then $y\in X\sm A.$ This proves $A=\{y\in X|0\leq y_1 \leq \alpha^A\}.$

To complete the proof consider another set $B\in \mathcal{A}_\succsim.$ Then there is some $k\in\{1,2\}$ and $\alpha^B>0$ such that $B=\{x\in X|0\leq x_k\leq \alpha^B\}.$ It only remains to show that $k=1.$ Suppose on the contrary, $k\neq 1.$ Then $k=2$ and $B=\{x\in X|0\leq x_2\leq \alpha^B\}.$

Recall by Proposition \ref{propone} that since $A,B$ are both unhappy sets, either $A\subseteq B$ or $B\subseteq A.$ By (\ref{a}), both $A,B$ are non empty. First suppose $A\subseteq B.$ Consider any $z=(z_1,z_2)\in A.$ Since $A(z_1)=\mathbb{R}_+,$ we have $(z_1,y_2)\in A$ for any $y_2>\alpha^B,$ so that $(z_1,y_2)\notin B,$ a contradiction. Next suppose $B\subseteq A$ and consider any $z=(z_1,z_2)\in B.$ Then $B(z_2)=\{y_1\in X_1|(y_1,z_2)\in B\}=\mathbb{R}_+.$ Thus $(y_1,z_2)\in B$ for any $y_1>\alpha^A,$ so that $(y_1,z_2)\notin A,$ a contradiction. So we must have $k=1.$ \qed

\vspace{0.2cm}\noi {\bf Proof of Theorem \ref{cor3}} A lexicographic preference on $X=\mathbb{R}^2_+$ is strong monotone and mildly continuous and by Lemma \ref{leximia}, any closed unhappy set of such a preference satisfies IMIA.

To prove the converse, let $\succsim$ be a complete, transitive, strong monotone and mildly continuous preference relation on $X=\mathbb{R}^2_+$ for which any closed unhappy set satisfies IMIA. Consider any $x=(x_1,x_2)\in \mathbb{R}^2_+$ such that $x_1,x_2$ are both positive. Let $L(x)$ be the lower contour set of $x$ and $\ol{L}(x)=L(x)\cup \partial{L(x)}$ be its closure. By Corollary \ref{cor2}, $\ol{L}(x)$ is a closed unhappy set, so it belongs to the family $\mathcal{A}_{\succsim}$ given in (\ref{a}). By Lemma \ref{bound} we conclude there is $\kappa(x)>0$ and a unique $i^*\in \{1,2\}$ such that $\ol{L}(x)=\{y\in \mathbb{R}^2_+|0\leq y_{i^*}\leq \kappa(x)\}.$ Without loss of generality, let $i^*=1.$ Then for any $x$ where $x_1,x_2$ are both positive we have $\ol{L}(x)=\{y\in \mathbb{R}^2_+|0\leq y_1\leq \kappa(x)\}.$

In what follows we show $\kappa(x)=x_1.$ Note that if $\kappa(x)<x_1,$ then $x$ will be outside $\ol{L}(x),$ so we must have $\kappa(x)\geq x_1.$ If $\kappa(x)>x_1,$ we can construct $\wt{x}\in \mathbb{R}^2_+$ such that $\wt{x}_{1}=\kappa(x)>x_1$ and $\wt{x}_2>x_2.$ Observe that $\wt{x}\in \ol{L}(x)$ and any neighborhood of $\wt{x}$ contains a point $y$ such that $y_1>\kappa(x),$ so that $y\notin \ol{L}(x).$ This shows $\wt{x}\in \partial{\ol{L}(x)}.$

Noting that $\wt{x}>x,$ by monotonicity we have $\wt{x}\succ x.$ Since $\succsim$ is mildly continuous and $\wt{x}\td x,$ there exists a neighborhood $B_\ve(\wt{x})$ such that all points there is strictly preferred to $x.$ So we have $\wt{x}\notin \partial{L(x)}.$ But we know $\wt{x}\in \partial{\ol{L}(x)}.$ This is a contradiction since $\partial{\ol{L}(x)}\subseteq \partial{L(x)}$ (for any set $A,$ the boundary of its closure is a subset of $\partial{A},$ see Chapter 3 of Mendelson 1990). This shows we must have $\kappa(x)=x_1.$ So for any $x$ where $x_1,x_2$ are both positive:
\be\label{low}\ol{L}(x)=\{y\in \mathbb{R}^2_+|0\leq y_1\leq x_1\}\ee To show that $\succsim$ is lexicographic, consider any $y,z\in \mathbb{R}^2_+$ such that $y\neq z.$ If $z_1>y_1,$ then $\exists$ $x$ with $x_1,x_2$ both positive such that $z_1>x_1>y_1.$ Then by (\ref{low}) it follows that $y\in \ol{L}(x)$ and $z\notin \ol{L}(x).$ Since $\ol{L}(x)$ is an unhappy set, we must have $z\succ y.$ This shows whenever $z_1>y_1,$ we must have $z\succ y.$ Finally let $y\neq z$ such that $z_1=y_1.$ Then by strong monotonicity, $z\succ y$ if $z_2>y_2$ and $y\succ z$ if $y_2>z_2.$ This shows that $\succsim$ is the lexicographic preference $1\succ^L 2.$ \qed

\vspace{0.2cm}\noi {\bf Proof of parts (ii)-(iii) of Lemma \ref{lex1}} Part (ii): Suppose $i\succ^* j$ and $j\succ^*k.$ Let $S=\{i,k\}.$ Since $\succsim$ is pairwise lexicographic, by Theorem \ref{tpair} it satisfies Axiom \ref{mon}, so $\succsim_S$ is strong monotone. 

Consider any $x^S=(x_i,x_k),$ $y^S=(y_i,y_k).$ To prove the `if' part of (\ref{pair1}), we have to show that if either (a) $[x_i>y_i]$ or (b) $[x_i=y_i, x_k>y_k]$ holds, then $(x^S,0^{N\sm S})\succ (y^S,0^{N\sm S}).$ If (b) holds, then $x^S\geq y^S$ and the result is immediate by the strong monotonicity of $\succsim_S.$ 

So suppose $x_i>y_i$ and let $T=\{i,j,k\}.$ Take any $\wt{x}_j>0$ and let 
$$x^{T}=(x_i,0,x_k),\wt{x}^{T}=(y_i,\wt{x}_j,x_k),y^{T}=(y_i,0,y_k)\mbox{ and }$$
$$x=(x^{T},0^{N\sm T}),\wt{x}=(\wt{x}^{T},0^{N\sm T}),y=(y^{T},0^{N\sm T})$$Note that $x=(x^S,0^{N\sm S})$ and $y=(y^S,0^{N\sm S}).$ Next observe that $x_i>\wt{x}_i=y_i$ and $x_{\ell}=\wt{x}_{\ell}$ for all $\ell\neq i,j.$ Since $i\succ^* j,$ by part (i) of Lemma \ref{lex1}, we have $x\succ \wt{x}.$ Also note that $\wt{x}_j>y_j=0$ and $\wt{x}_{\ell}=y_{\ell}$ for all $\ell\neq j,k.$ Since $j\succ^* k,$ again by part (i) of Lemma \ref{lex1}, we have $\wt{x}\succ y.$ Thus $x\succ\wt{x}\succ y.$ By transitivity of $\succsim,$ we have $x=(x^S,0^{N\sm S})\succ y=(y^S,0^{N\sm S}),$ which proves the `if' part of (\ref{pair1}).  

To prove the `only if' part of (\ref{pair1}), suppose $(x^S,0^{N\sm S})\succ (y^S,0^{N\sm S}).$ Then we cannot have $[x_i=y_i, x_k=y_k].$ So, if neither (a) $[x_i>y_i],$ nor (b) $[x_i=y_i, x_k>y_k]$ holds, then we must have either 
$[y_i>x_i]$ or $[y_i=x_i,y_k>x_k],$ but in that case by the same reasoning as in the proof of the `if' part we shall have  $(y^S,0^{N\sm S})\succ (x^S,0^{N\sm S}),$ a contradiction. This shows that the `only if' part of (\ref{pair1}) also holds and therefore $i\succ ^*k.$  

\vspace{0.1cm} Part (iii): Since for any $i,j\in N,$ either $i\succ^*j$ or $j\succ^*i,$ the result follows from part (ii) by applying  Observation \ref{obs1}. \qed 

\begin{ob}\label{obs1} {\it Consider a finite set $A$ that has $n\geq 2$ objects. Let $R$ be a relation on $A$ such that} (i) {\it for any two different $x,y\in A,$ either $xRy$ or $yRx$ but not both and} (ii) {\it for any $x,y,z\in A,$ if $xRy$ and $yRz,$ then $xRz.$ Then the objects of $A$ can be ordered as $a_1Ra_2R\ldots a_{n-1}R a_n$ and we can write $A=\{a_1,\ldots,a_n\}.$} 

\end{ob}

\noi {\bf Proof} We prove the observation by induction on $n.$ The result clearly holds for $n=2.$ For $n\geq 3,$ suppose the result is true for any finite set that has $m\leq n-1$ objects. Consider a set $A$ that has $n$ objects and fix a specific object $x\in A.$ Let $S=\{a\in A|aRx\}$ and $T=\{a\in A|xRa\}.$ By property (i) of $R,$ it follows that $S,T$ are disjoint and $S\cup T=A\sm \{x\}.$ Let $|S|=s,$ $|T|=t.$ Then $0\leq s,t\leq n-1.$ Since properties (i), (ii) hold for both $S,T,$ by induction hypothesis, we have $S=\{a_1,\ldots,a_s\}$ such that $a_1R\ldots Ra_s$ and $T=\{b_1,\ldots,b_t\}$ such that $b_1R \ldots Rb_t.$ Since $a_sRx$ and $xRb_1,$ by property (ii) of $R,$ we have $a_1R\ldots Ra_sRxRb_1R \ldots Rb_t.$ \qed

\section*{References}

\noi Arrow KJ (1951) {\it Social Choice and Individual Values.} (Wiley, New York; 2nd ed., 1963).

\vspace{0.1cm} \noi Bahrampour M, Byrnes J, Norman R, Scuffham PA, Downes M (2020) Discrete choice experiments to generate utility values for multi-attribute utility instruments: A systematic review of methods. {\it European Journal of Health Economics} 21(7):983-992. 

\vspace{0.1cm} \noi Bettman JR, Luce MF, Payne JW (1998) Constructive consumer choice processes. {\it Journal of Consumer Research} 25(3):187-217.

\vspace{0.1cm} \noi Birnbaum MH, Gutierrez RJ (2007) Testing for intransitivity of preferences predicted by a lexicographic semi-order. {\it Organizational Behavior and Human Decision Processes} 104(1):96-112.

\vspace{0.1cm} \noi Bossert W, Pattanaik PK, Xu Y (1994) Ranking opportunity sets: An axiomatic approach. {\it Journal of Economic Theory} 63(2):326-345.

\vspace{0.1cm}\noi Campbell D, Hutchinson WG, Scarpa R (2006) Lexicographic preferences in discrete choice experiments: consequences on individual-specific willingness to pay estimates. Sustainability Indicators and Environmental Valuation Working Papers 12224, Fondazione Eni Enrico Mattei (FEEM).

\vspace{0.1cm}\noi Chistyakov V, Chumakova K (2018) Restoring indifference classes via ordinal numbers under the discrete leximin and leximax preference orderings. {\it Journal of the New Economic Association} 39(3):12-31.

\vspace{0.1cm} \noi Debreu G (1954) Representation of a preference ordering by a numerical function. Thrall R, Coombs C, Davis R eds. {\it Decision Processes} (Wiley, New York), 159-165.

\vspace{0.1cm}\noi Fishburn PC (1975) Axioms for lexicographic preferences. {\it Review of Economic Studies} 42(3):415-419.

\vspace{0.1cm}\noi Ford JK, Schmitt N, Schechtman SL, Hults BH, Doherty ML (1989) Process tracing methods: Contributions, problems, and neglected research questions. {\it Organizational Behavior and Decision Processes} 43(1):75-117.

\vspace{0.1cm}\noi Geanakoplos J (2005) Three brief proofs of Arrow's impossibility theorem. {\it Economic Theory} 26(1):211-215.

\vspace{0.1cm}\noi Gigerenzer G, Goldstein DG (1999) Betting on one good reason: The take the best heuristic. Gigerenzer G, Todd PM, ABC Research Group eds. {\it Simple Heuristics that Make Us Smart} (Oxford University Press),  75-95.

\vspace{0.1cm}\noi Gigerenzer G, Gaissmaier W (2011) Heuristic decision making. {\it Annual Review of  Psychology} 62(1):451-482.

\vspace{0.1cm}\noi  Gilbride T, Allenby GM (2004) A choice model with conjunctive, disjunctive, and compensatory screening rules. {\it Marketing Science} 23(3):391-406.

\vspace{0.1cm}\noi Gonz\'{a}lez-Vallejo C (2002) Making trade-offs: A probabilistic and context-sensitive model of
choice behavior. {\it Psychological Review} 109(1):137-155.

\vspace{0.1cm}\noi Hogarth RM, Karelaia N (2006) Regions of rationality: Maps for bounded agents. {\it Decision Analysis} 3(3):124-144.

\vspace{0.1cm}\noi Katsikopoulos KV (2011) Psychological heuristics for making inferences: Definition, performance, and the emerging theory and practice. {\it Decision Analysis} 8(1):10-29.

\vspace{0.1cm}\noi Katsikopoulos KV (2013) Why do simple heuristics perform well in choices with binary attributes? {\it Decision Analysis} 10(4):327-340. 

\vspace{0.1cm}\noi Keeney RL, Raiffa H (1993) {\it Decisions with Multiple Objectives}. Cambridge University Press. 

\vspace{0.1cm}\noi Kurz-Milcke E, Gigerenzer G (2007) Heuristic decision making. {\it Marketing: Journal of Research and Management} 3(1):48-56.

\vspace{0.1cm}\noi Lancsara E, Louviere J (2006) Deleting ‘irrational’ responses from discrete choice experiments: a case of investigating or imposing preferences? {\it Health Economics} 15(8):797-811.

\vspace{0.1cm}\noi Leong W, Hensher DA (2012) Embedding multiple heuristics into choice models: An exploratory analysis. {\it Journal of Choice Modelling} 5(3):131-144.

\vspace{0.1cm}\noi McIntosh E, Ryan M (2002) Using discrete choice experiments to derive welfare estimates for the provision of elective surgery: Implications of discontinuous preferences. {\it Journal of Economic Psychology} 23(3):367-382.

\vspace{0.1cm}\noi Meenakshi JV, Banerji A, Manyong V, Tomlins K, Mittal N, Hamukwala P (2012) Using a discrete choice experiment to elicit the demand for a nutritious food: Willingness-to-pay for orange maize in rural Zambia. {\it Journal of Health Economics} 31(1):62-71.

\vspace{0.1cm}\noi Mendelson B (1990) {\it Introduction to Topology.} Third Edition, Dover Publications.

\vspace{0.1cm}\noi Mitra M, Sen D (2014) An alternative proof of Fishburn's axiomatization of lexicographic preferences. {\it Economics Letters} 124(2): 168-170.

\vspace{0.1cm}\noi Mitra M, Sen D (2014) Subsistence, saturation and irrelevance in preferences. MPRA Paper No. 60614.

\vspace{0.1cm}\noi Petri H, Voorneveld M (2016) Characterizing lexicographic preferences. {\it Journal of Mathematical Economics} 63(C):54-61.

\vspace{0.1cm}\noi Rubinstein A (2012) {\it Lecture Notes in Microeconomic Theory: The Economic Agent.} Second Edition,  Princeton University Press.

\vspace{0.1cm}\noi Rudin W (1976) {\it Principles of Mathematical Analysis.} McGraw-Hill Inc.

\vspace{0.1cm}\noi S{\ae}lensminde K (2002) The impact of choice inconsistencies in stated choice studies. {\it Environmental and Resource Economics} 23(4):403-420.

\vspace{0.1cm}\noi Scott A (2002) Identifying and analysing dominant preferences in discrete choice experiments:
An application in health care. {\it Journal of Economic Psychology} 23(3):383-398.

\vspace{0.1cm}\noi Tversky A (1969) Intransitivity of preferences. {\it Psychological Review} 76(1):31-48.

\vspace{0.1cm}\noi Tversky A (1972) Elimination by aspects: A theory of choice. {\it Psychological Review} 79(4):281-299.

\vspace{0.1cm}\noi Yee M, Dahan E, Hauser J, Orlin J (2007) Greedoid-based noncompensatory inference. {\it Marketing Science} 26(4):532-549.

\end{document}